\documentclass[aps,prb,a4paper,10pt,twocolumn,amsmath,showpacs,superscriptaddress]{revtex4-1} 
\usepackage[hmargin=2cm,vmargin=2.5cm]{geometry}

\usepackage[pdftex]{color}
\usepackage{verbatim}
\pdfoutput=1

\usepackage{booktabs,microtype,afterpage} 

\usepackage[colorlinks,plainpages=false,linkcolor=blue,urlcolor=blue,citecolor=blue,pdfpagemode=UseNone,pdfstartview=FitBH]{hyperref}
\usepackage[charter,greekuppercase=italicized]{mathdesign}
\definecolor{red}{rgb}{0.85,.1,0}
\definecolor{green}{rgb}{0.0,0.6,0.0}
\definecolor{orange}{rgb}{1,0.5,0}
\usepackage{graphicx}    
\graphicspath{{./}{figure/}{/home/d748/Partage/Dati/Ge_2016/Giovam/z_papero/figure/nafeas_images/}}
\DeclareGraphicsExtensions{.eps,.png,.pdf,.jpg}

\usepackage{threeparttable}
\usepackage[referable]{threeparttablex}

\newcommand{\ypis}{Yb$_2$\-Pd$_{2}$\-In$_{1-x}$\-Sn$_{x}$}
\newcommand{\yps}{Yb$_2$\-Pd$_{2}$\-Sn}

\newcommand{\ypi}{Yb$_2$\-Pd$_{2}$\-In}

\newcommand{\musr}{$\mu$SR}

\begin{document}

\title{Pressure-Induced Antiferromagnetic Dome in the Heavy-Fermion \texorpdfstring{\ypis}{Yb2Pd2In(1-x)Snx} System} 

\author{G.~Lamura}
\email[Corresponding author: ]{gianrico.lamura@spin.cnr.it}
\affiliation{CNR-SPIN, Corso Perrone 24, 16152 Genova, Italy}
\author{I.~J.~{Onuorah}}
\affiliation{Department of Mathematical, Physical and Computer Sciences, University of Parma, 43124 Parma, Italy}
\author{P.~{Bonf\`{a}}}
\affiliation{Department of Mathematical, Physical and Computer Sciences, University of Parma, 43124 Parma, Italy}
\affiliation{Centro S3, CNR-Istituto Nanoscienze, 41125 Modena, Italy}

\author{S.~Sanna}
\affiliation{Department of Physics and Astronomy, University of Bologna, 40127 Bologna, Italy}
\author{Z.~{Shermadini}}
\affiliation{Laboratory for Muon Spin Spectroscopy, Paul Scherrer Institut (PSI), Villigen, CH-5232 Villigen, Switzerland}
\author{R.~{Khasanov}}
\affiliation{Laboratory for Muon Spin Spectroscopy, Paul Scherrer Institut (PSI), Villigen, CH-5232 Villigen, Switzerland}
\author{J.-C.~Orain}
\affiliation{Laboratory for Muon Spin Spectroscopy, Paul Scherrer Institut (PSI), Villigen, CH-5232 Villigen, Switzerland}
\author{C.~{Baines}}
\affiliation{Laboratory for Muon Spin Spectroscopy, Paul Scherrer Institut (PSI), Villigen, CH-5232 Villigen, Switzerland}
\author{F.~Gastaldo}
\affiliation{Dipartimento di Chimica e Chimica Industriale, University of Genova, 16146 Genova, Italy}
\author{M.~Giovannini}
\affiliation{Dipartimento di Chimica e Chimica Industriale, University of Genova, 16146 Genova, Italy}
\author{I.~\u{C}url\'{\i}k}
\affiliation{Faculty of Humanities and Natural Sciences, University of Pre\v{s}ov, SK 081 16 Pre\v{s}ov, Slovakia}
\author{A.~Dzubinska}
\affiliation{CPM-TIP, University Pavol Jozef Safarik, 041 54  Kosice, Slovakia}
\author{G.~Pristas}
\affiliation{Institute of  Experimental Physics of the Slovak Academy of Sciences (IEP SAS), 040 01 Ko\v{s}ice, Slovakia}
\author{M.~Reiffers}
\affiliation{Faculty of Humanities and Natural Sciences, University of Pre\v{s}ov, SK 081 16 Pre\v{s}ov, Slovakia}
\affiliation{Institute of  Experimental Physics of the Slovak Academy of Sciences (IEP SAS), 040 01 Ko\v{s}ice, Slovakia}
\author{A.~Martinelli}
\affiliation{CNR-SPIN, Corso Perrone 24, 16152 Genova, Italy}
\author{C.~Ritter}
\affiliation{Institut Laue-Langevin, 38042 Grenoble, France}
\author{B.~Joseph}
\affiliation{GdR IISc-ICTP, Elettra-Sincrotrone, Basovizza, 34149 Trieste, Italy}
\author{E.~Bauer}
\affiliation{Institute of Solid State Physics, TU Wien, A-1040 Wien, Austria}
\author{R.~{De~Renzi}}
\affiliation{Department of Mathematical, Physical and Computer Sciences, University of Parma, 43124 Parma, Italy}
\author{T.~Shiroka}
\affiliation{Laboratory for Muon Spin Spectroscopy, Paul Scherrer Institut (PSI), Villigen, CH-5232 Villigen, Switzerland}
\affiliation{Laboratorium f\"ur Festk\"orperphysik, ETH-H\"onggerberg, CH-8093 Z\"urich, Switzerland}
%
%
%
\begin{abstract}
In the heavy-fermion system \ypis, the interplay of crystal-field splitting, 
Kondo effect, and Ruderman-Kittel-Kasuya-Yosida interactions leads to complex 
chemical-, pressure-, and magnetic-field phase diagrams, still to be explored 
in full detail. By using a series of techniques, we show that even modest 
changes of parameters other than temperature are sufficient to induce 
multiple quantum-critical transitions in this highly susceptible 
heavy-fermion family. In particular, we show that, above $\sim 10$\,kbar, hydrostatic pressure not only induces an antiferromagnetic phase at 
low temperature, but it likely leads 
to a reorientation of the Yb magnetic moments and/or the competition among different antiferromagnetic configurations. 
\end{abstract}
\maketitle
\section{Introduction} 
Yb-based intermetallic compounds represent one of the most interesting 
classes of materials for investigating the interplay of crystal-field 
splitting, Kondo effect, and Ruderman-Kittel-Kasuya-Yosida (RKKY) interactions. 
Given their comparable energy scales, it is not surprising that the 
resulting ground state can easily be tuned via external pressure, chemical substitution, or 
applied magnetic fields \cite{Gegenwart2002,Gegenwart2008}. 
Within the larger class of heavy-fermion (HF) compounds, Yb-based materials 
are among the best examples of quantum critical systems showing a 
non-Fermi liquid behavior. This is due to 
the manifest sensitivity of the Yb$^{2+}$ electronic configuration to chemical- and external pressure. 
Thus, an increase in pressure ``squeezes'' one of the Yb $4f$-electrons out of its shell, driving the non magnetic Yb$^{2+}$ ($4f^{14}$, $J=0$) 
to magnetic Yb$^{3+}$ ($4f^{13}$, $J=7/2$,  $\mu=4.52~\mu_\mathrm{B}$). 
As a consequence, the Kondo effect is weakened, while the RKKY-mediated exchange 
interactions among the Yb ions are enhanced, a favorable condition for the onset of a 
long-range magnetically ordered phase \cite{Bauer2004, Bauer2005, Bauer2010, Muramatsu2011, Yamaoka17}. 
This is precisely the case of \ypis, previously investigated via 
macroscopic techniques and lately studied also via $\mu$SR, at both ambient- and high-pressure conditions, 
in the stoichiometric $x=1$ case \cite{Muramatsu2011}. While, in general, there is consensus 
with regard to  
the broad picture, key details are still missing. In particular, the 
joint effects of chemical- and applied pressure, as well as the resulting phase diagrams and the type of magnetic order remain largely unexplored to date.\\
Here, by combining a range of experimental techniques, including magnetic susceptibility, muon-spin 
rotation ($\mu$SR), x-ray, and neutron diffraction, we map out 
the temperature-, pressure-, and composition phase diagram of this prototypical HF system. 
In particular, we show that externally applied pressure reinforces the effects of 
chemical pressure (here achieved via In/Sn substitution), 
by extending the magnetically ordered dome and by reordering the Yb magnetic moments into a new antiferromagnetic phase, whose critical temperature reaches $T_\mathrm{N}\sim$ 4.9 K at 23.4 kbar at the optimum doping.

Numerical calculations and 
symmetry considerations proved essential in clarifying 
the new magnetic structure adopted at high pressure.
\section{Crystal structure, x-ray diffraction and dc-magnetization}
A series of \ypis\ polycrystalline samples with nominal compositions 
$x = 0$, 0.3, 0.6, 0.8 were prepared from stoichiometric amounts of 
pure elements by high-frequency melting of the constituent materials 
in a closed tantalum crucible. A subsequent one-week heat treatment 
at 1250\,K was used to ensure chemical homogeneity. 
All the compounds of this series crystallize in the tetragonal $P4/mbm$ 
space group, where Yb occupies the $4h$ sites. Systematic x-ray diffraction and dc magnetization measurements performed on all 
the samples confirmed the absence of spurious phases (within the sensitivity 
of the respective techniques \cite{Rx}).
Samples were further investigated via high-pressure synchrotron 
x-ray diffraction at the Elettra source and neutron powder diffraction at the Institute Laue-Langevin. The detailed results of the above measurements are reported in the Apps.~\ref{app:A}--\ref{app:C} and in Ref.~\cite{Mauro2018}. 
\begin{figure}[tbh]
	\centering
	\includegraphics[width=0.4\textwidth]{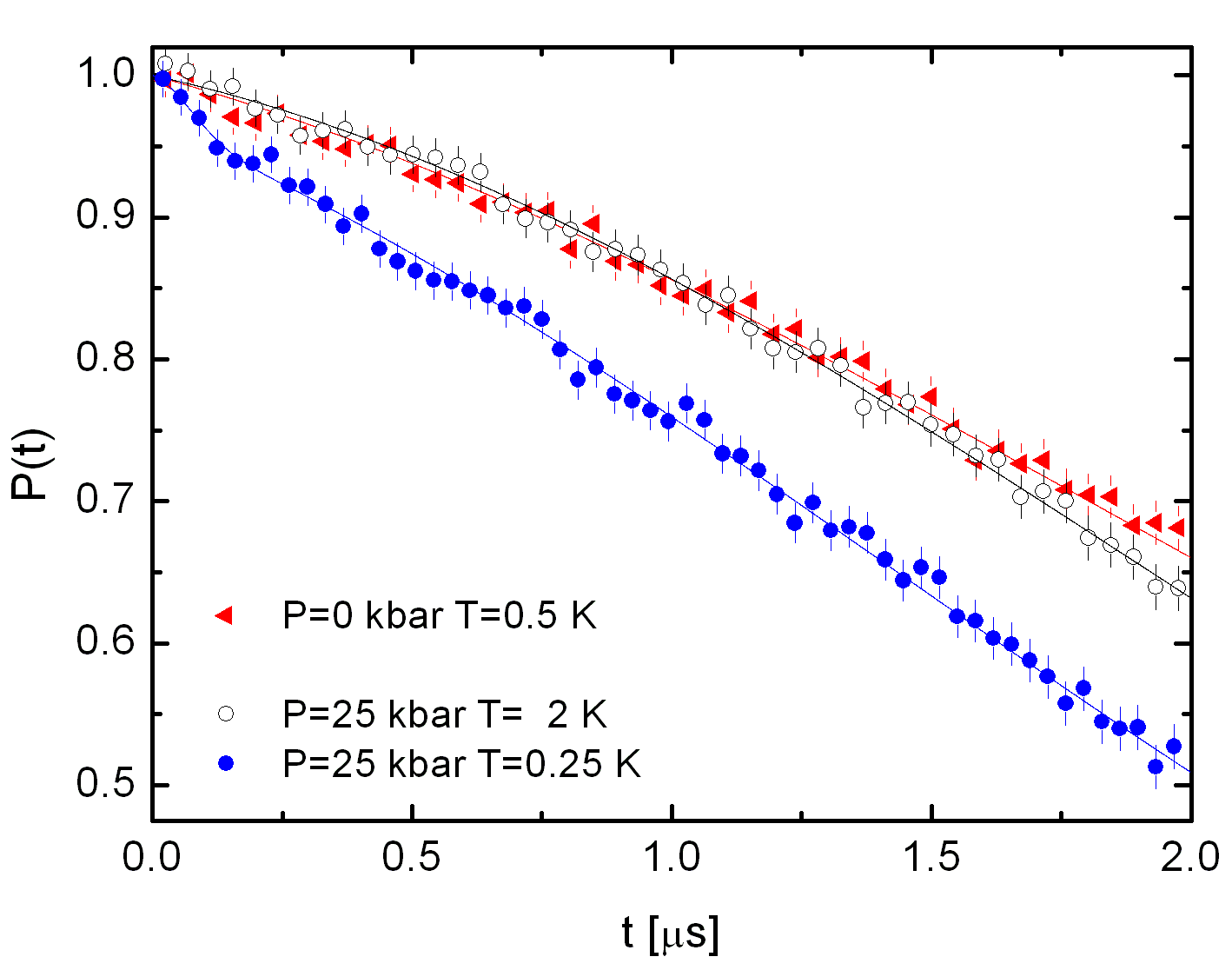} 
	\caption{\label{fig:asym00}Time-domain ZF-\musr\ polarization for 
		the $x=0$ case measured at $T=0.5$\,K for $p=0$\,kbar (\textcolor{red}{\large $\blacktriangleleft$}) 
		and at $T=2$\,K ({\Large $\circ$}) and 0.25\,K (\textcolor{blue}{\Large $\bullet$}) 
		for $p=25$\,kbar. Only  in the last case a tiny depolarization 
		is seen at short times ($t < 0.2$\,$\mu$s).}
\end{figure}
\section{\label{GPD} \musr\ measurements}
The muon-spin relaxation measurements under hy\-dro\-sta\-tic\--pres\-sure conditions 
in the 0--25\,kbar range, from 0.25 to 10\,K, were carried out at the General Purpose Decay-Channel (GPD) spectrometer of the S$\mu$S muon source of the Paul Scherrer Institut (Villigen, Switzerland).  The external pressure was applied by  using a double-wall piston-cylinder pressure cell made of MP35N alloy. Daphne oil 7373 was used as a pressure-transmitting medium to 
achieve nearly hydrostatic conditions across the whole pressure range 
\cite{KhasanovHPR,ShermadiniHPR}. To determine the exact pressure at low temperature, 
a small piece of indium was placed next to the sample. The pressure-dependent 
shift of its superconducting transition $T_c(p)$ was determined via 
ac susceptometry \cite{KhasanovHPR,ShermadiniHPR}. 
\begin{figure*}[tbh]
	\centering
	\includegraphics[width=0.3225\textwidth]{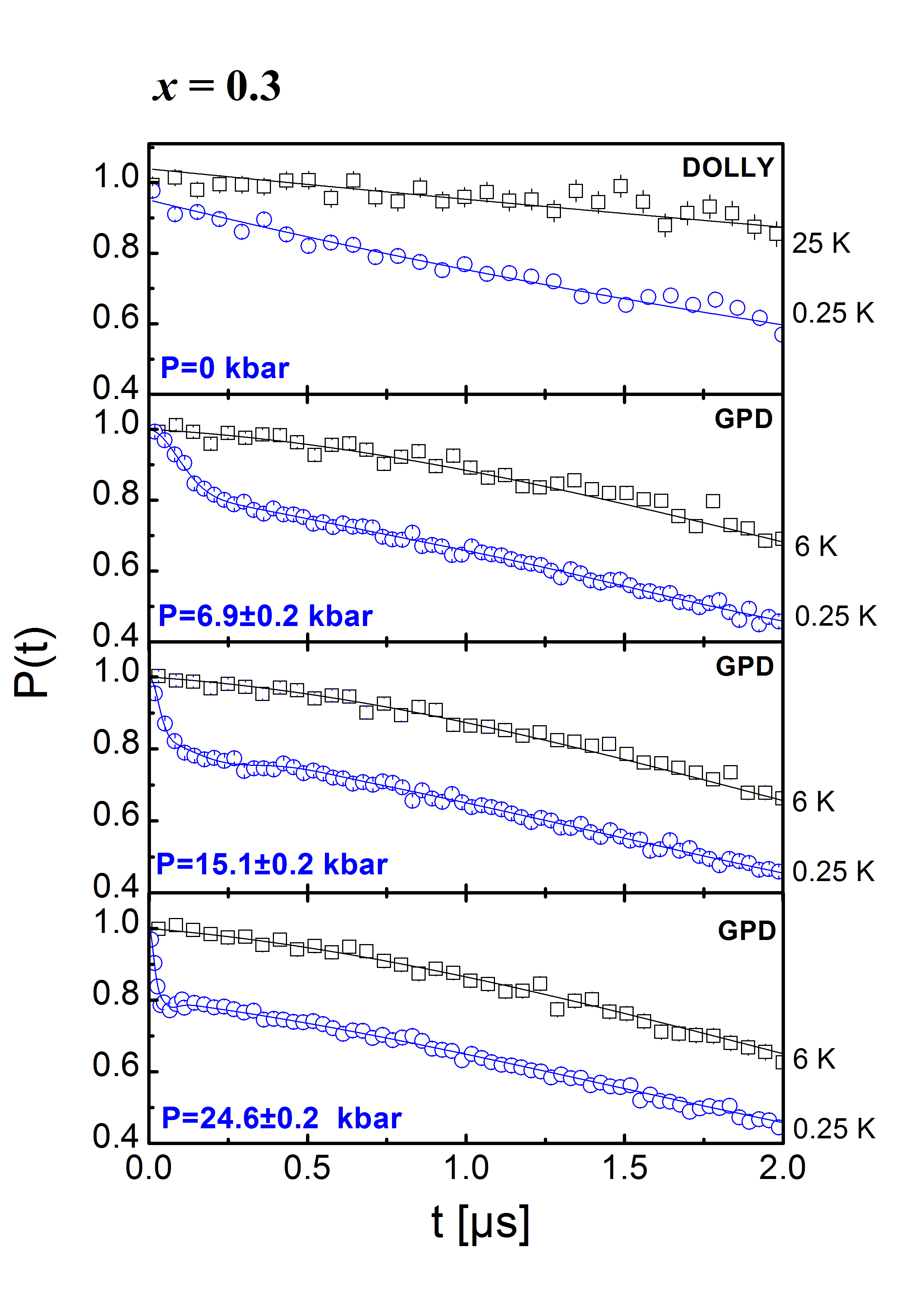} 
	\includegraphics[width=0.3235\textwidth]{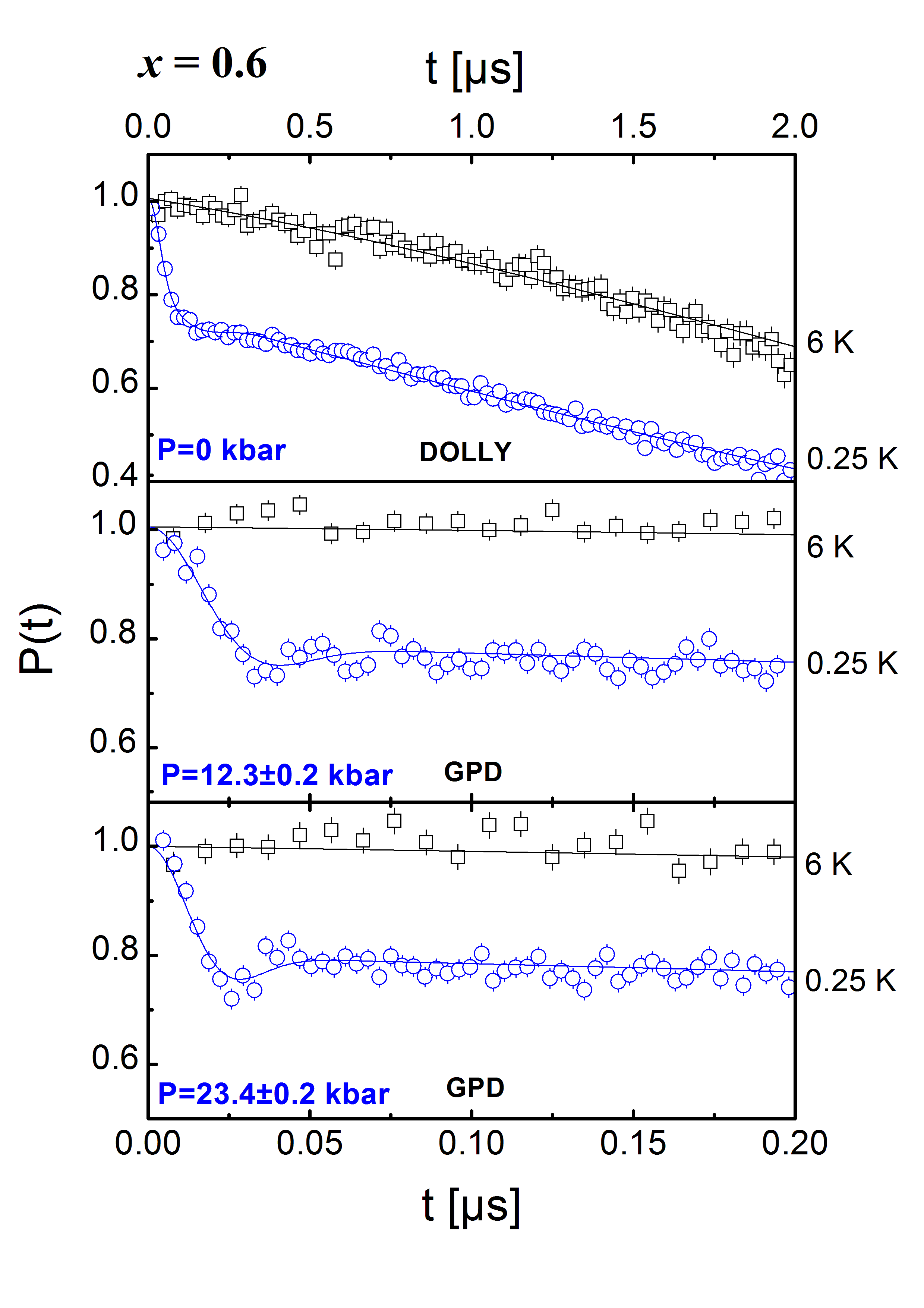} 
	\includegraphics[width=0.3315\textwidth]{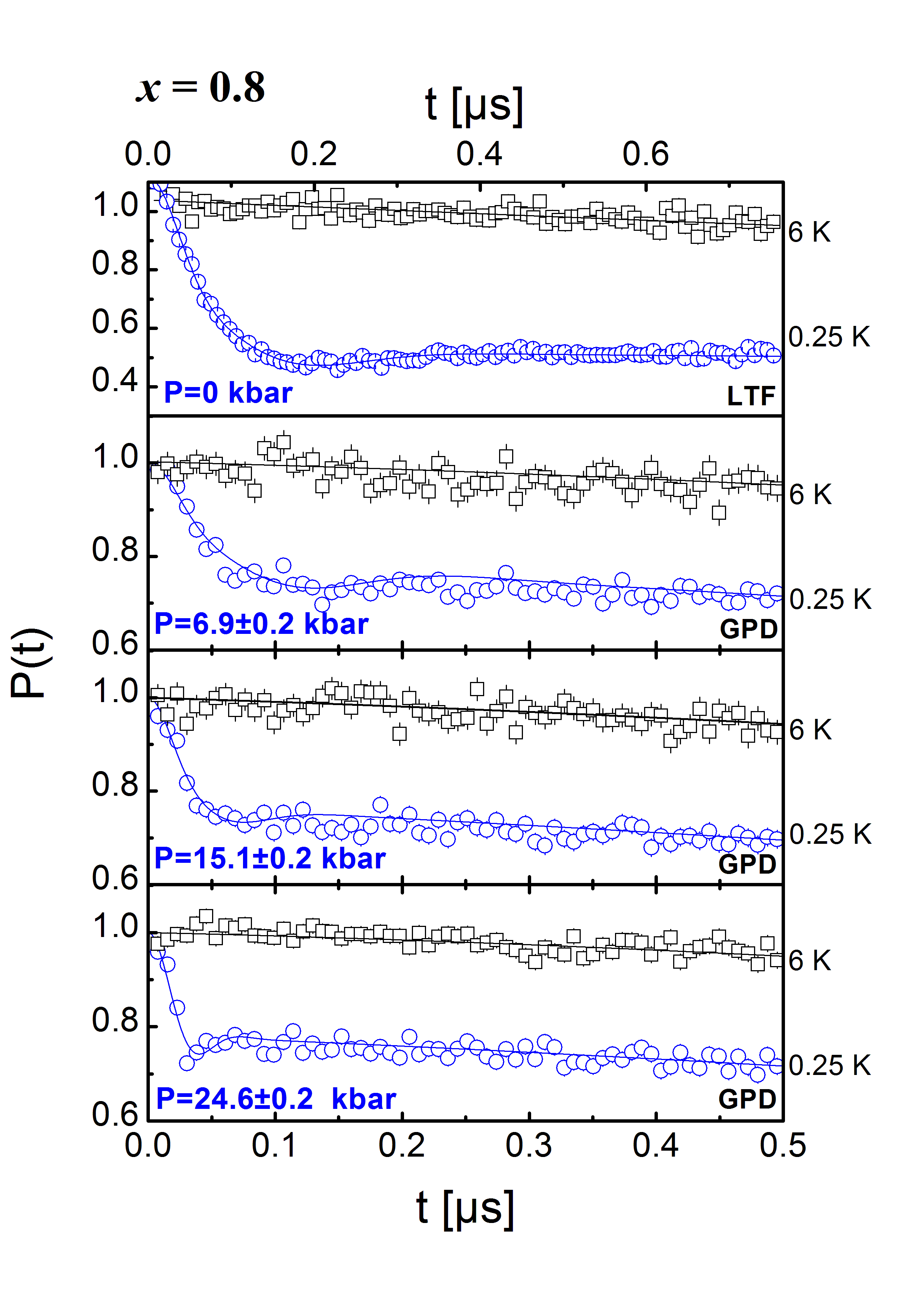} 
	\caption{\label{fig:asym}Time-domain ZF-\musr\ polarization 
		measured at base temperature (0.25\,K) and above $T_\mathrm{N}$ 
		(typically at 6\,K) at ambient- and under applied-pressure conditions 
		for the $x > 0$ case. 
		From left to right, the Sn concentrations are $x = 0.3$, 0.6, and 0.8. The appearance of oscillations under applied pressure indicates the onset of a pressure-induced antiferromagnetic phase.}
\end{figure*}
Due to the small mass of In, its contribution to the \musr\ background is negligible. 

Figures~\ref{fig:asym00} and \ref{fig:asym} show the time-dependent, 
zero-field (ZF) muon-spin depolarization at short-time for all 
the samples under test at representative temperatures and applied pressures. As a general feature, highly-damped coherent oscillations are seen to develop upon increasing the Sn content or the applied pressure, 
thus providing key evidence about the onset of a long-range magnetic order with a large degree of inhomogeneity. To determine the parameters of these coherent precessions and to disentangle the spurious contribution of the pressure cell  
to the total signal, 
the time-dependent muon-spin depolarization was fitted by using the model:
\begin{equation}
\begin{split}
\label{eq:spin_prec}
P(t)=\frac{A^{\mathrm{ZF}}(t)}{A_{\mathrm{tot}}^{\mathrm{ZF}}(0)} = 
a_\mathrm{bg}\, g(t) +\left[1-a_\mathrm{bg}\right] \cdot\\ 
\sum_{i=1}^{N} w_i \cdot \left[a_{T_i} \, f_i(\gamma_{\mu}B_{\mu} t) \, D_{T_i}(t) + a_{L_i} \, D_{L_i}(t) \right].
\end{split}
\end{equation}
Here $A_{\mathrm{tot}}^{\mathrm{ZF}}(0)$ is the high-temperature value of the initial asymmetry, whereas $a_\mathrm{bg}$ accounts for the 
fraction of incoming muons stopped outside the sample. Regarding this last parameter, three cases can be distinguished: \textit{i}) $a_\mathrm{bg}$ 
is almost zero during the reference experiments performed at the low-background spectrometer Dolly (samples with $x=0.3$ and 0.6); \textit{ii}) $a_\mathrm{bg}$ coincides with the muon 
fraction implanted in the Ag sample holder during the experiments on $x=0.8$ sample at Low Temperature Facility (LTF) at the SµS, PSI. In this case we assume $g(t)=e^{-\lambda_\mathrm{Ag}t}$, with both $a_\mathrm{bg}$ 
and $\lambda_\mathrm{Ag}$ being determined at the lowest temperature 
and kept fixed during subsequent fits \cite{background}. 
\textit{iii}) During the GPD measurements, $a_\mathrm{bg}$ accounts for the fraction of muons implanted in the pressure cell \cite{TF_mu}. In this case, $g(t)$ represents a Gaussian Kubo-Toyabe function multiplied by an exponential damping \cite{KhasanovHPR,ZF_mu}. 

The coherent muon precession in the magnetically\--or\-der\-ed phase is described by the $f(t)$ function, while the $D_{T_i}(t)$ and $D_{L_i}(t)$ 
terms account for a possible damping. Here $D_{T_i}(t)$ reflects the static distribution of local magnetic fields, whereas $D_{L_i}(t)$ describes the dynamical relaxation processes. $B^{i}_\mu$ is the magnetic 
field at the $i$-th muon implantation site and $\gamma_{\mu} = 2\pi \times 135.53$\,MHz/T is the muon gy\-ro\-mag\-ne\-tic ratio; $a_{T_i}$ and $a_{L_i}$ refer to 
muons probing local magnetic fields in the transverse (T) or longitudinal (L) direction with respect to the initial mu\-on\--spin polarization. The sum over $i$ generalizes Eq.~(\ref{eq:spin_prec}) to the case of  
diverse inequivalent crystallographic implantation sites, whose population weights $w_i$ satisfy the normalization condition $\sum_i^{N} w_i =1$. In the following, we specify  
the above parameters in the context of each case.

\begin{figure*}[tbh]
	\centering  
	\includegraphics[width=0.32\textwidth]{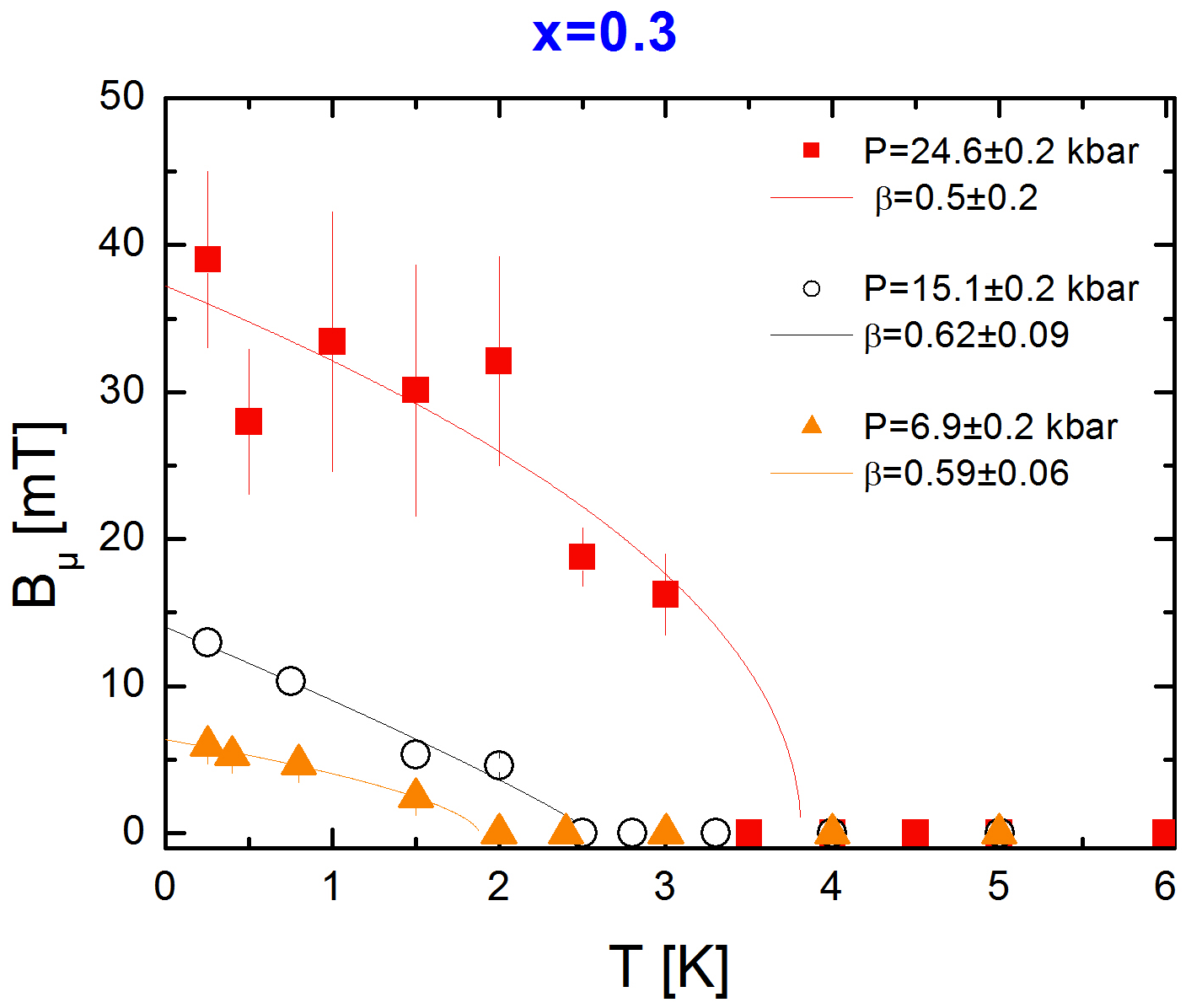} 
	\llap{\parbox[b]{14mm}{\large{(a)}}}
	\includegraphics[width=0.32\textwidth]{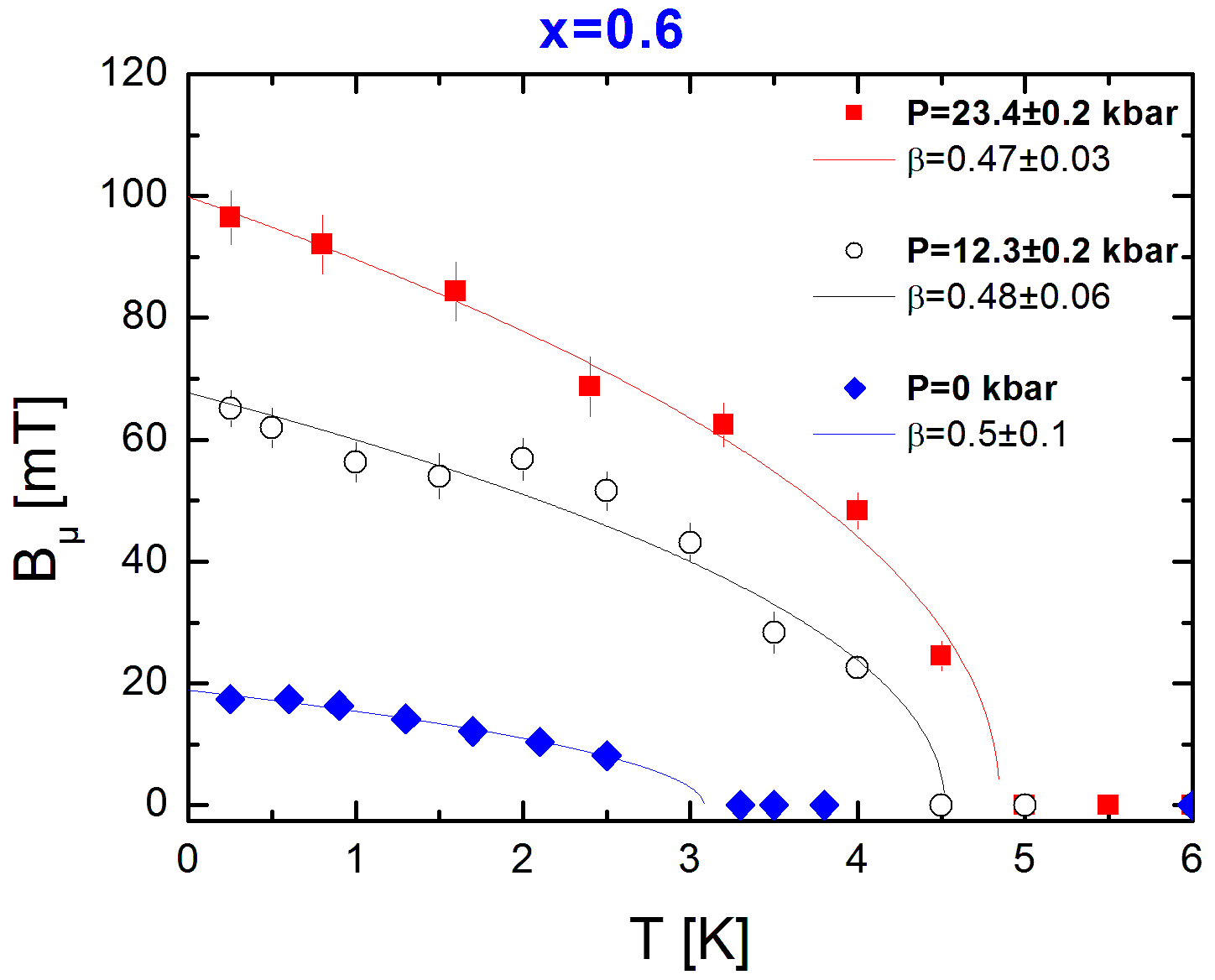}
	\llap{\parbox[b]{14mm}{\large{(b)}}}	
	\includegraphics[width=0.32\textwidth]{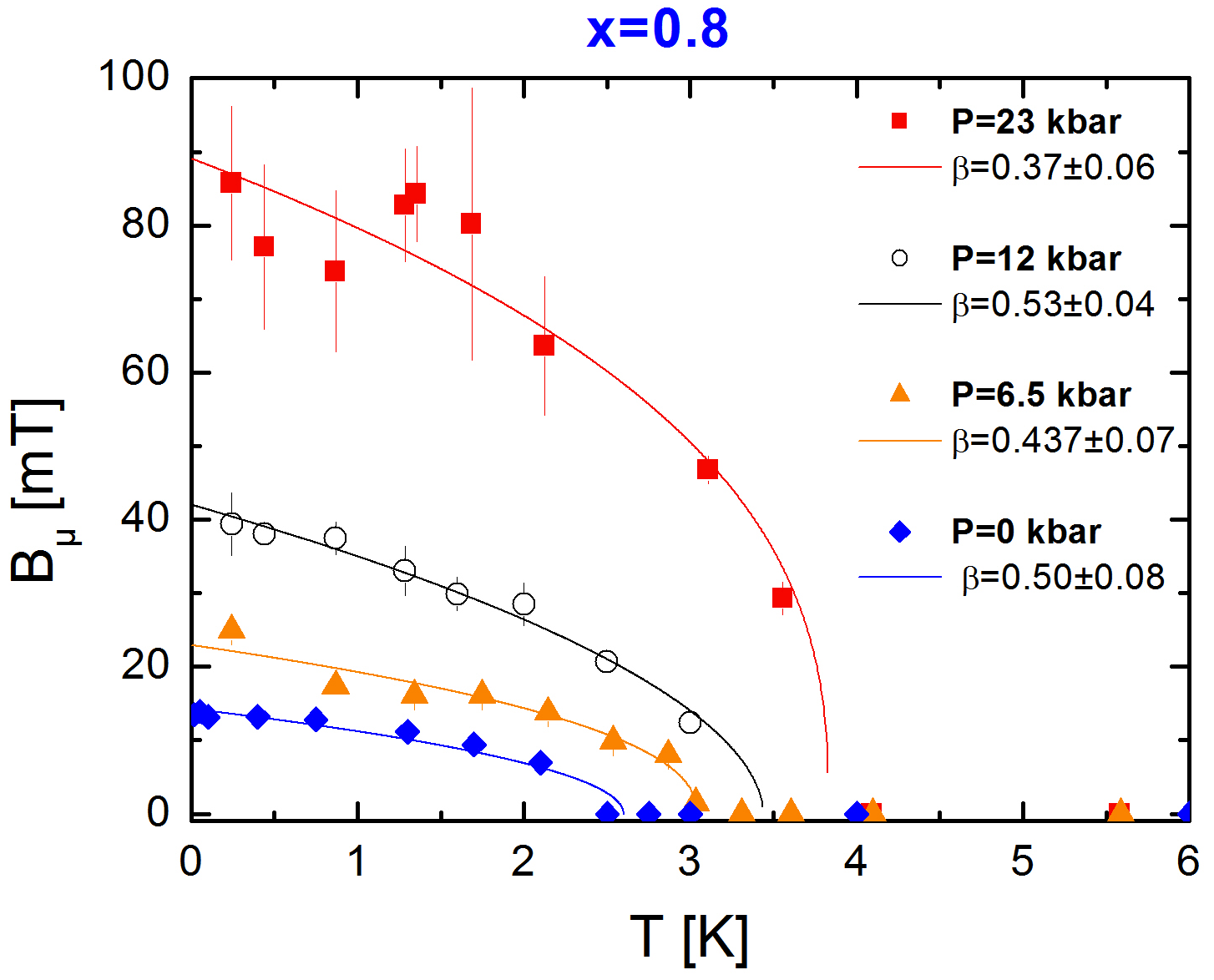}
	\llap{\parbox[b]{14mm}{\large{(c)}}}
	\\[\baselineskip]
	\includegraphics[width=0.32\textwidth]{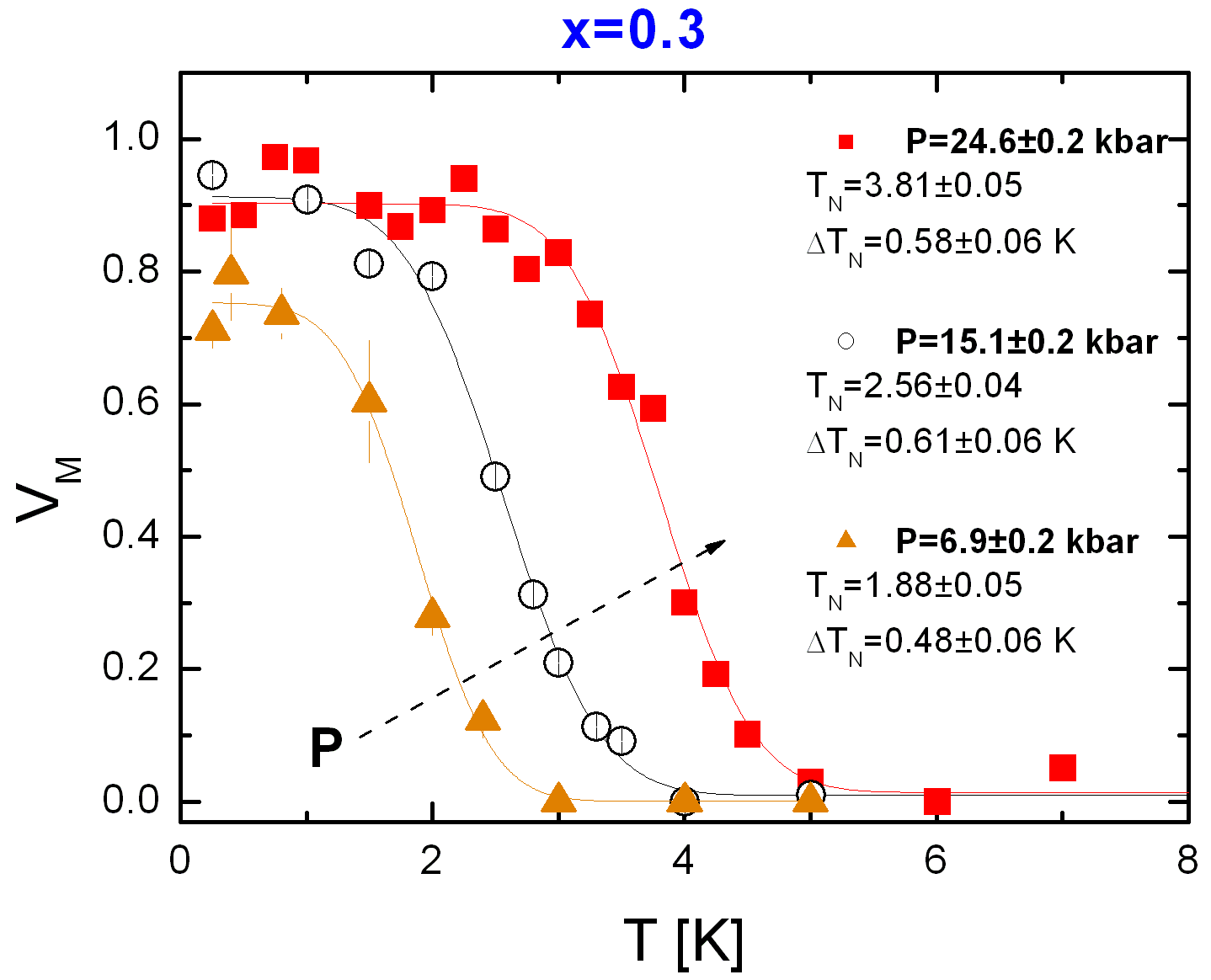}
	\llap{\parbox[b]{14mm}{\large{(d)}}}
	\includegraphics[width=0.32\textwidth]{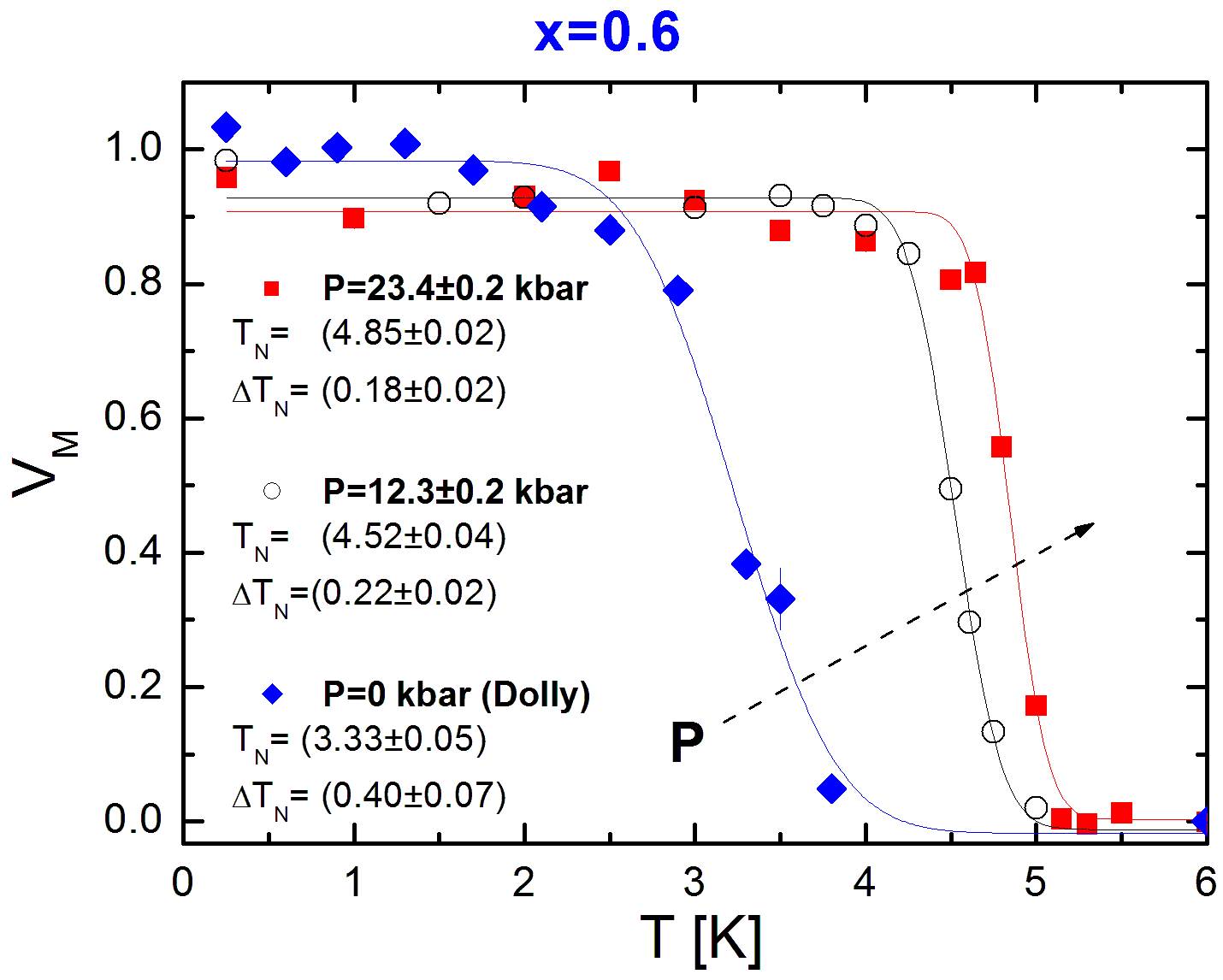}
	\llap{\parbox[b]{14mm}{\large{(e)}}}
	\includegraphics[width=0.32\textwidth]{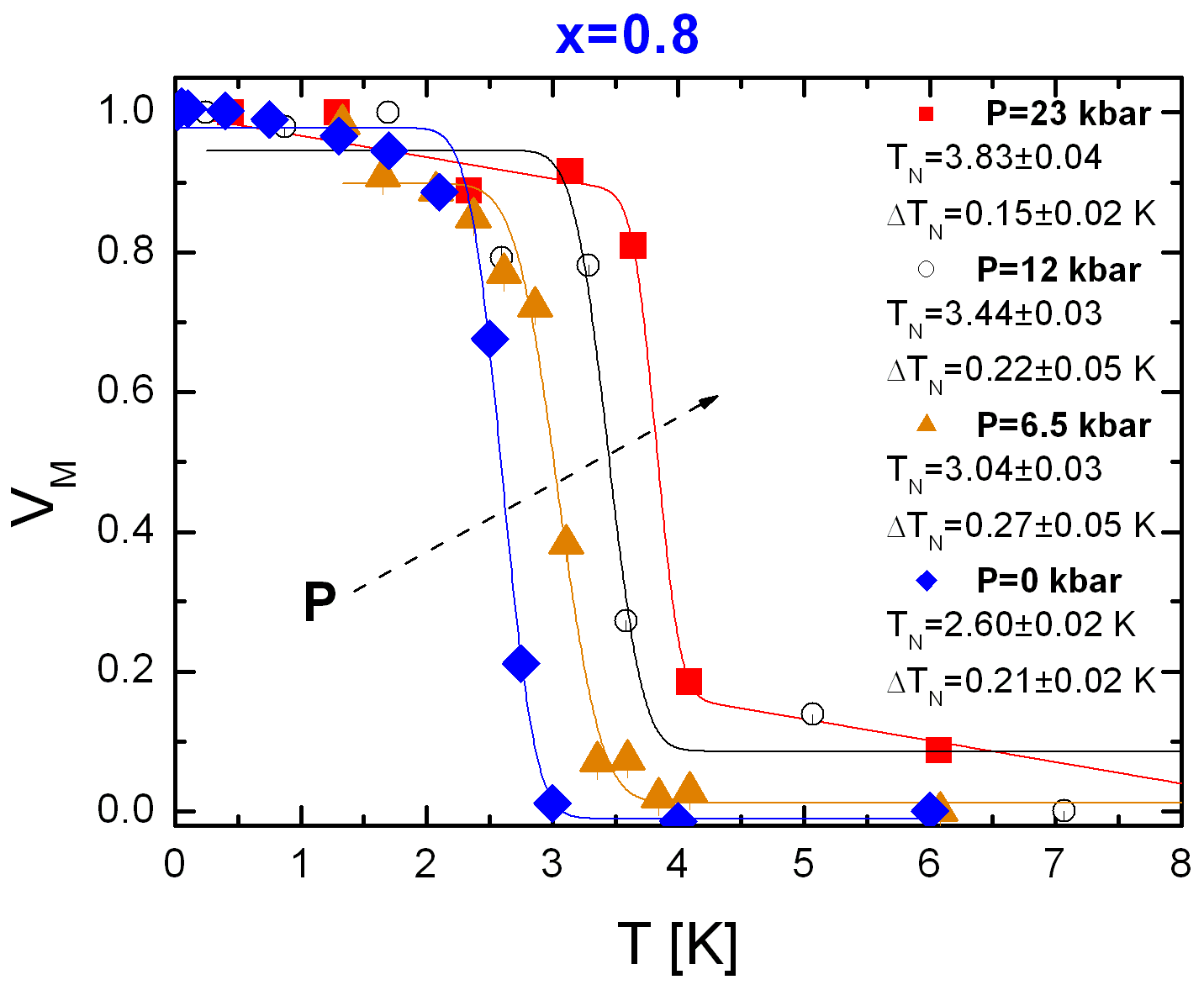} 
	\llap{\parbox[b]{14mm}{\large{(f)}}}
	\caption{\label{fig:Bmu_Vmag}Internal magnetic fields $B_{\mu}$ (top 
		panels) and magnetic volume fractions (bottom panels), as determined  
		from fits of the ZF- and wTF $\mu$SR data collected at different applied pressures, for the $x=0.3$, 0.6, and 0.8 case. The continuous lines in the top panels represent numerical fits according to a Landau mean field theory $B=B_0\cdot(1-T/T_N)^{\beta}$ with beta as free parameter ($\beta=0.5$ in the case of 2nd order transitions). In the bottom panels, the continuous lines represent numerical fits to an \textit{erf} model function (see text for details).
	}
\end{figure*}

\paragraph{$x=0$.}
Figure~\ref{fig:asym00} shows the short-time zero-field (ZF) \musr\ 
depolarization for the pure In case. The polarization $P(t)$ 
can be fitted by a single Gaussian decay at both standard- and 
applied-pressure conditions, provided the temperature is not very low. This type of decay is normally associated with randomly-oriented nuclear dipolar fields. However, when (at 25\,kbar) the temperature is lowered to 0.25\,K, a small transverse component appears, indicating 
that $\sim$15\% of the sample develops weak magnetic correlations. Such behavior suggests that \ypi\ is on the verge of quantum criticality, with a magnetic phase transition presumably occurring at higher pressure.

\paragraph{$x=0.3$.}
The low-background ambient-pressure data taken at 
$T = 25$\,K [Fig.~\ref{fig:asym}(a), top panel] are well fitted 
by a single Gaussian depolarization, reflecting the randomly oriented nuclear dipolar fields. At the lowest temperature a small depolarization appears, suggesting that also in this case, at ambient pressure the system is likely at the verge of a magnetic instability. Interestingly, by progressively increasing the pressure (to 12.3 and 23.4 kbar) a strongly damped oscillation appears at 0.25 K. In this case, the fitting function describing the sample contribution consists of two transverse components, one of which is a damped Gaussian cosine \cite{cos} and the other is represented by a Gaussian decay. The corresponding longitudinal components can be merged into a single Lorentzian decay. 
It is worth noting that the frequency of the oscillating part increases rapidly with increasing pressure, suggesting the proximity to a pressure-induced magnetic order.
\paragraph{$x=0.6$.}
As shown in Fig.~\ref{fig:asym} (top panel), the low background data taken at ambient pressure and $T = 0.25$\,K are well fitted by one transverse component, consisting of a damped Gaussian cosine \cite{cos} and a single Lorentzian decay. This differs from the $x = 0.3$ case where, at ambient pressure, a weakly decaying Gaussian function is sufficient 
to describe the data. The previous two-component fitting function was successfully adopted also for the measurements under pressure [Fig.~\ref{fig:asym}(b), lower 
panels]. Here, too, a sudden increase of muon-precession frequency with increasing pressure occurs.
\paragraph{$x=0.8$.} 
At ambient pressure, the low background data taken at $T = 6$\,K, i.e., above $T_\mathrm{N}$, [Fig.~\ref{fig:asym}, top panel] 
are well fitted by a single Gaussian depolarization, reflecting the randomly oriented nuclear dipolar moments. At the lowest temperature, the damped oscillation could be fitted by two transverse components, of which one is a damped Gaussian cosine term \cite{cos} and the other is a Gaussian decay. The two longitudinal signals, again merged into one, are fitted by a single Lorentzian exponential. Upon applying pressure, the same fitting function was adopted [Fig.~\ref{fig:asym}, lower panels]. Also in this case, the frequency of the oscillating component increases rapidly with increasing pressure.\\

Panels~\ref{fig:Bmu_Vmag}(a)-(c) show the evolution of the oscillating component of the local field $B_\mu$ as a function of temperature for all the applied pressures. Generally, in all cases, we note a steep increase of the internal magnetic field. In particular, for $x=0.6$, the internal field shows a fivefold enhancement from its ambient-pressure value.

\begin{figure}[tbh]
	\includegraphics[width=0.5\textwidth]{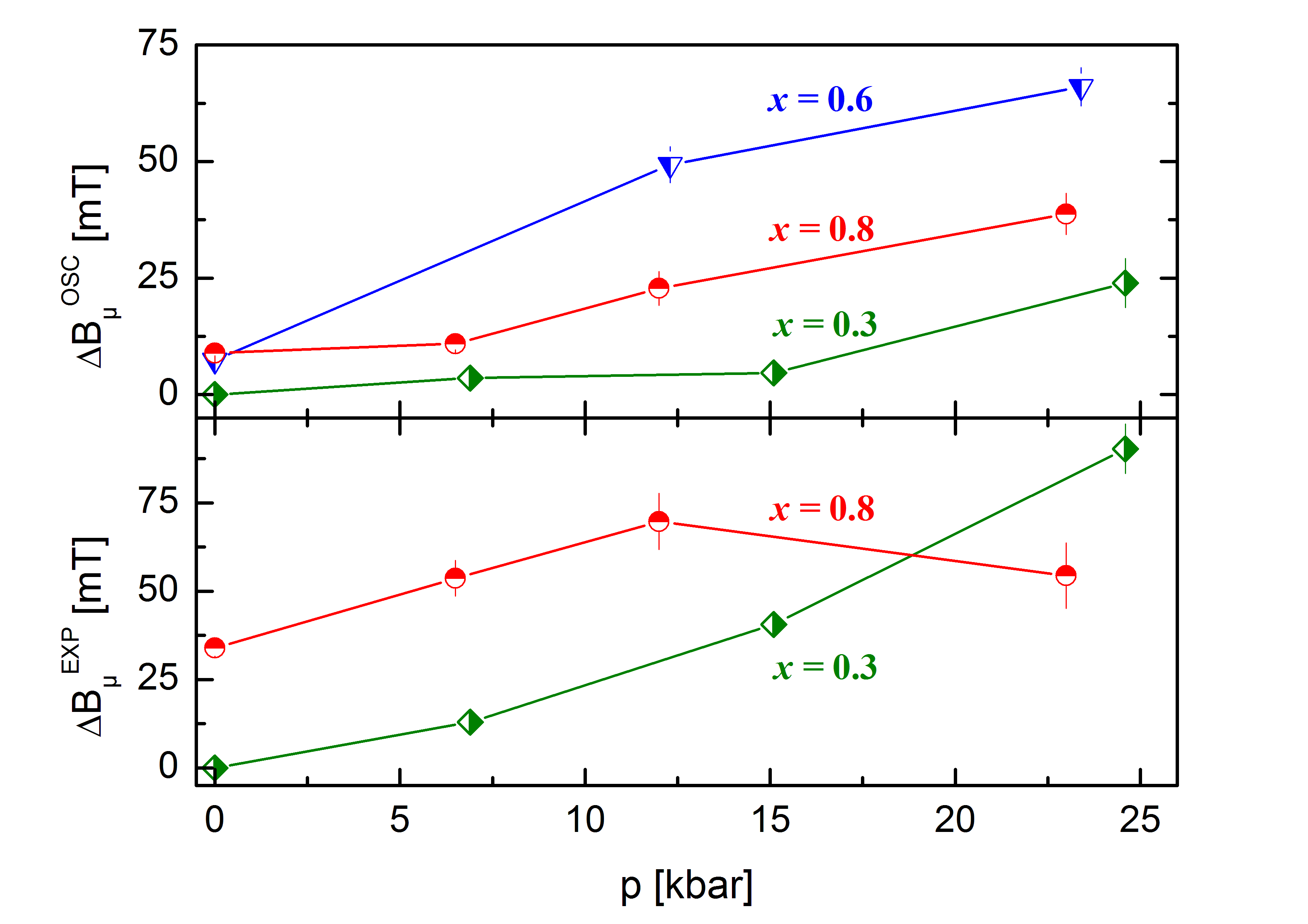} 
	\caption{\label{fig:Delta_Bmu}Internal magnetic-field widths for the 
	two transverse components, $\Delta B_{\mu}^\mathrm{osc}$ (top panel) and $\Delta B_{\mu}^\mathrm{exp}$ (bottom panel), measured at the lowest temperature for the samples with 
	$x = 0.3$, 0.6, and 0.8. The field width for the non oscillating Gaussian decay could not be detected for $x = 0.6$ (see text for details).}
\end{figure}

Figure~\ref{fig:Delta_Bmu} summarizes the evolution of the local-field widths for both transverse components at the lowest temperature as a function of applied pressure. We note a remarkable broadening of both widths at the muon implantation sites upon increasing pressure.

The magnetic volume fraction $V_M$ in the ordered phase was determined in two different ways. In case of ambient pressure experiments, it was extracted from the total longitudinal component by means of $V_M(T) = 3(1-a_{\parallel}^\mathrm{tot.})/2$ \cite{Shiroka2011,AsymFrac}.
In case of applied pressure, the magnetic volume fraction was estimated from the temperature evolution of the oscillating paramagnetic fraction in a weak transverse field (wTF) \musr\ experiment ($\mu_0H=3$\,mT). 
The evolution of the magnetic volume fraction with the applied pressure is shown in Fig.~\ref{fig:Bmu_Vmag}(d)-(f) for all those cases where a magnetically ordered phase could be detected. 

We note two important features: \textit{i}) the transition widths remain rather narrow, never exceeding 1\,K, irrespective of the applied pressure; \textit{ii}) the magnetic ordering temperature increases steeply with 
applied pressure. For instance, in the optimally substituted $x = 0.6$ sample, $T_\mathrm{N}$ at 23\,kbar reaches twice its ambient-pressure value. 
In the explored pressure range $T_\mathrm{N}$ increases almost linearly with pressure. The differential increment (slope) is maximal for $x=0.3$ (ca.\ 0.15\,K/kbar), then it gradually decreases, to saturate at $0.07$\,K/kbar at higher tin content. Such values are remarkably high 
for a heavy-fermion compound (compared, for instance, with 0.016\,K/kbar for CeNiSiH$_x$ \cite{Isnard2016}).

\begin{figure*}[tbh]
	\centering
	\includegraphics[width=0.65\textwidth]{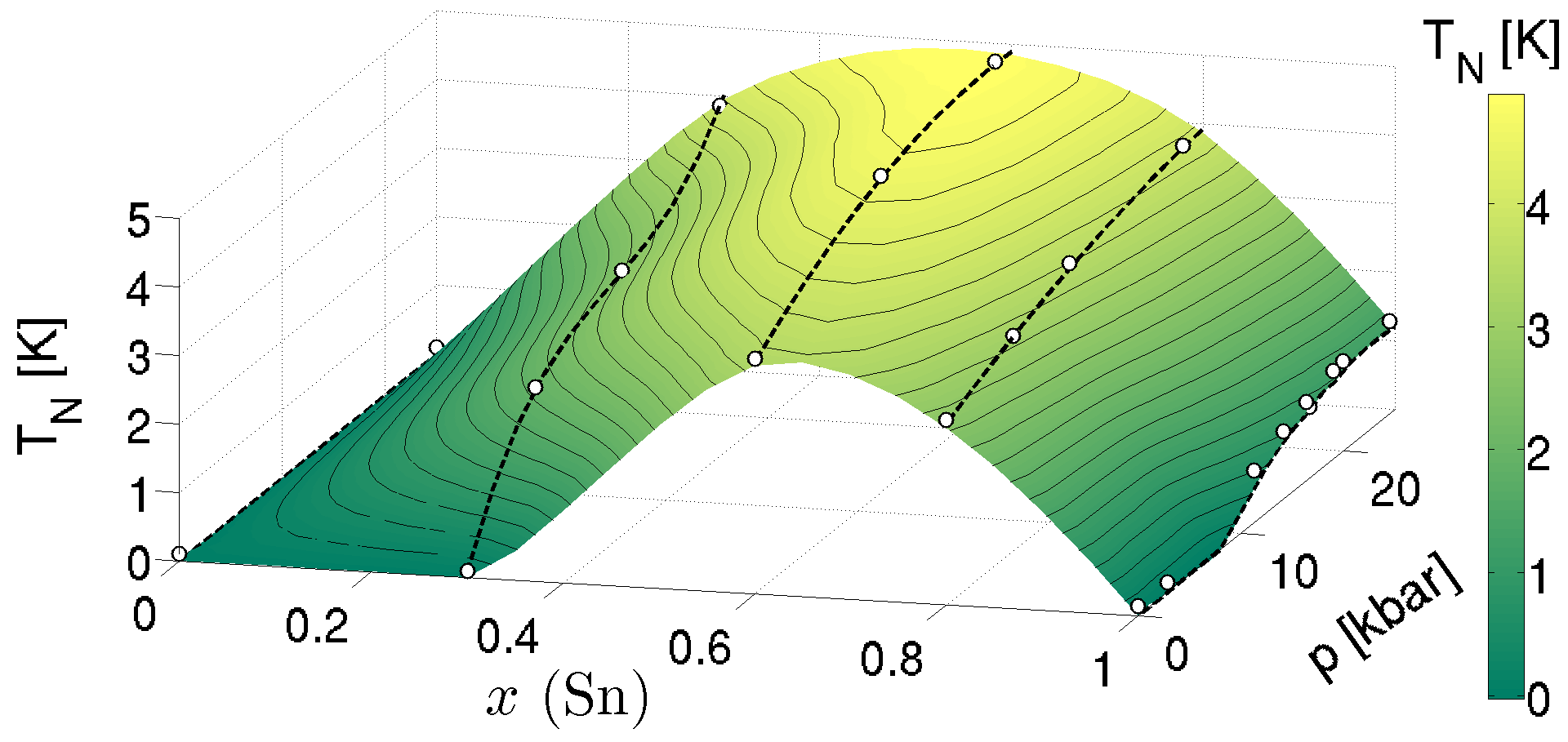}
    \llap{\parbox[b]{0mm}{\large{(a)}}}
    \includegraphics[width=0.68\textwidth]{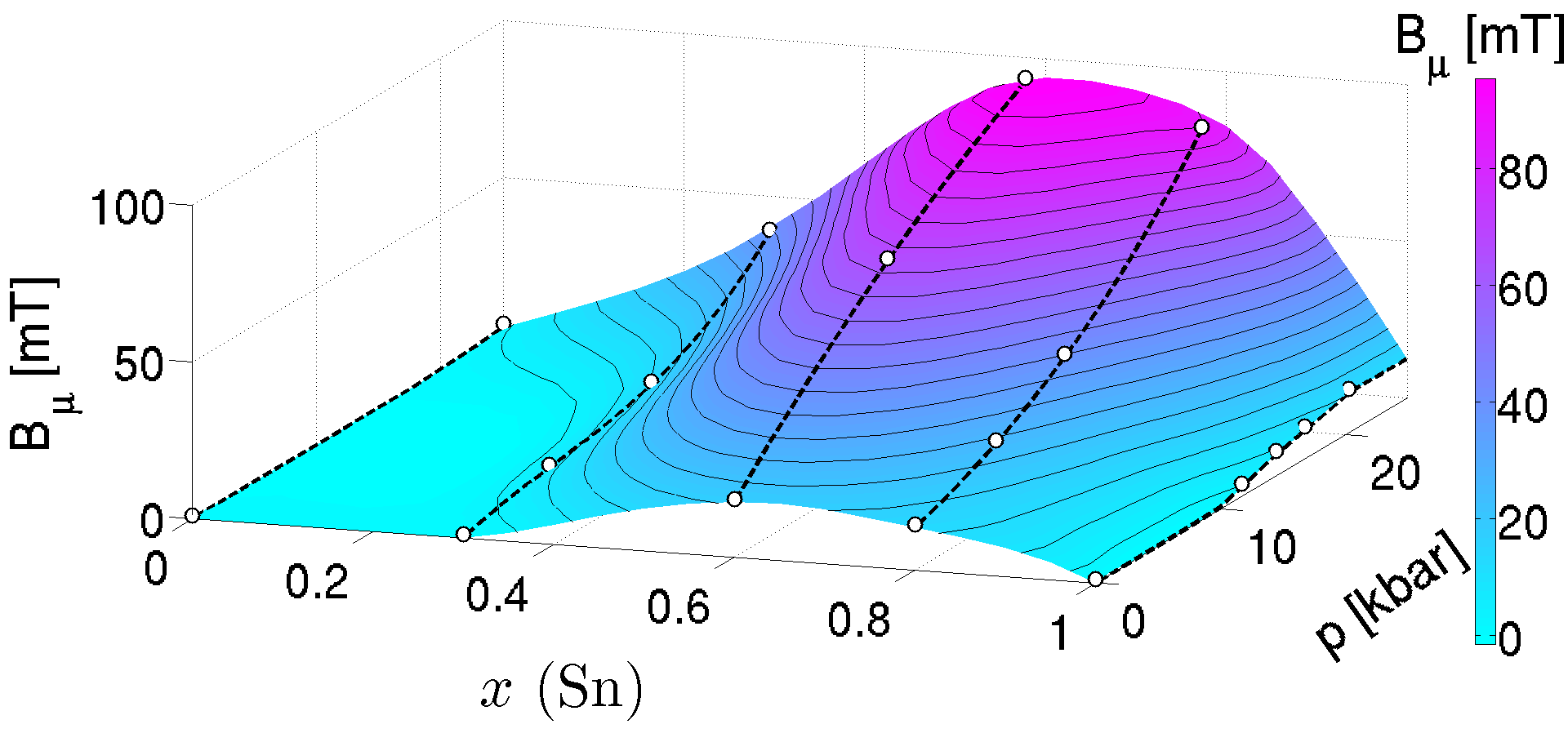}
	\llap{\parbox[b]{0mm}{\large{(b)}}}
	\caption{\label{fig:3D_diag}$p$-$T$-$x$ (a) and $p$-$B$-$x$ (b) phase 
	diagrams of the \ypis\ system. White dots represent experimental points, 
	while 3D surfaces and dashed lines are interpolations  
	obtained via polynomial fits. Note the increased $B_{\mu}$ field at high pressure.} 
\end{figure*}

The datasets shown in Fig.~\ref{fig:Bmu_Vmag} allow us to build the full
$p$-$T$-$x$ and $p$-$B_{\mu}$-$x$ phase diagrams, as shown in Fig.~\ref{fig:3D_diag}. Some interesting features can be highlighted: \textit{i}) For $x=0$, at $T=0.25$\,K and $p=25$\,kbar the system is already on the verge of quantum 
criticality, as almost 15\% of the sample develops weak magnetic 
correlations, suggesting that a possible phase transition might occur at higher pressures. 
\textit{ii}) For $x=0.3$, at $T=0.25$\,K and ambient pressure the system 
is again at the verge of a magnetic instability, with only a small 
volume fraction developing magnetic correlations. Upon increasing pressure, 
a long-range magnetic order appears, with transition temperatures and local magnetic 
fields reaching 3.8\,K and 40\,mT, respectively, at 24.6\,kbar. 
\textit{iii}) In the $x=0.6$ (optimally doped) and $x=0.8$ cases a long-range magnetic order is already present at ambient pressure 
and $T = 0.25$\,K. At the highest applied pressure the magnetic 
ordering temperature almost doubles and the internal field shows 
a fivefold increase from its ambient-pressure value. \textit{iv}) Upon increasing pressure, we observe  a remarkable broadening of the field width at each muon implantation site. \textit{v}) For $x=1.0$, previous experiments  
have shown that a magnetically 
ordered phase appears only for pressures above 10\,kbar \cite{Muramatsu2011}.\\ \\
In summarizing this section we conclude that: (a) below $T_\mathrm{N}$, all the tested compounds order magnetically over the whole sample volume, (b) the magnetic transition widths 
remain narrow even at the highest applied pressures, while $T_\mathrm{N}$ 
increases steeply with pressure; (c) the internal field increases 
steeply above a critical pressure $p_\mathrm{cr} \sim 10$\,kbar. This is accompanied by a progressive broadening of the field widths. 
\begin{table*} [!tbp]
 \begin{threeparttable}
\caption{\label{tab:table1}The ten magnetic structures calculated by MAXMAGN code (labeled I to X) 
for the parent space group $P4/mbm$, that allow a nonzero 
magnetic moment on Yb, with propagation vector (0,0,0.5) \cite{Bauer2010, Mauro2018}. 
We report also the magnetic structure on the four symmetry-equivalent 
magnetic Yb ions in half of the unit cell, doubled along the $c$ axis 
and the calculated muon dipolar field $B_\mathrm{dip}$ in units of mT.~\tnotex{tnxx62}}
\begin{ruledtabular}
 \begin{tabular}{ c @{\hspace{1em}} l @{\hspace{1.2em}} l @{\hspace{1.2em}} l @{\hspace{1.2em}} l @{\hspace{1.2em}}l  l} 
     & \multicolumn{6}{c}{Magnetic structure~\tnotex{tnxx59}} \\
     \cline{3-6} \\
     Label & Magnetic group & Yb1 & Yb2& Yb3 &  Yb4 & {$B_\mathrm{dip}$\,[mT]} (s.d.)\\ 
     \colrule
    I    &$P_{c}4_{2}/mnm$    & ($m_x$, $-m_x$, 0)   & ($-m_x$, $m_x$, 0)    &  ($-m_x$, $-m_x$, 0)  & ($m_x$, $m_x$, 0)   & 6.18 (1.91)\\
    II   &$P_{c}4_{2}/mbc$    & ($m_x$, $m_x$, 0)    & ($-m_x$, $-m_x$, 0)     &  ($m_x$, $-m_x$, 0)  & ($-m_x$, $m_x$, 0)  & 3.79 (1.66)\\
    III  &$P_{c}4_{2}/mbc$    & (0, 0, $m_z$)        & (0, 0, $m_z$)          &  (0, 0, $-m_z$)       & (0, 0, $-m_z$)    & 3.58 (1.76)\\
    \textbf{IV}   &$P_{c}4/mnc$        & (0, 0, $m_z$)        & (0, 0, $m_z$)          &  (0, 0, $m_z$)        & (0, 0, $m_z$)     & 245.22 (4.95)\\
    V    &$P_{c}4/mnc$        & ($m_x$, $m_x$, 0)    & ($-m_x$, $-m_x$, 0)     & ($-m_x$, $m_x$, 0)   &($m_x$, $-m_x$, 0)  & 4.66 (1.68)\\
    VI   &$P_{c}4/mbm$        & ($m_x$,$-m_x$, 0)    & ($-m_x$, $m_x$, 0)       & ($m_x$, $m_x$, 0)   & ($-m_x$, $-m_x$, 0) & 6.87 (1.89)\\
    VII  &$C_{c}mcm$          & (0, 0, 0)            & (0, 0, 0)               & (0, 0, $m_z$)        &( 0, 0,$-m_z$)     & 2.25 (1.21)\\
    \textbf{VIII}~\tnotex{tnxx63} &$C_{c}mcm$            &($m_x$, $m_x$, 0)    & ($m_x$, $m_x$, 0)       &($m_y$,$m_y$, 0)     &($m_y$, $m_y$, 0) & 122.61 (4.53)\\
    IX   &$P_{b}nma$          & (0, 0, $m_z$)        & (0, 0, $-m_z$)          & (0, 0, $-m_z$)      &(0, 0, $m_z$)       & 3.03 (1.25)\\
    \textbf{X}~\tnotex{tnxx63}    &$P_{b}nma$            & ($m_x$, $m_y$, 0)     & ($m_x$, $m_y$, 0)      & ($-m_x$, $m_y$, 0)  &   ($-m_x$, $m_y$, 0) & 100.17 (4.54)\\
   
    \end{tabular} 
  \end{ruledtabular}
    \begin{tablenotes}
             \item[a]  \label{tnxx62} Dipolar field at the muon position (0.5, 0.5, 0.5) for $x=0.5$. 
             The In/Sn substitution-induced displacements on the Yb ions were considered. The values 
             in parenthesis labeled (s.d.) represent the standard deviation considering different random displacements. 
             \item[b]  \label{tnxx59}Yb1, Yb2, Yb3, Yb4 at (0.1724, 0.6724, $c/2$), (0.8276, 0.3276, $c/2$), (0.3276, 0.1724, $c/2$), (0.6724, 0.8276, $c/2$); $c=0.5$. 
             \item[c] \label{tnxx63} In these cases $m_x=m_y$ was considered. 
 \end{tablenotes} 
 \end{threeparttable}
  \
\end{table*}
\section{DFT and dipolar field calculations} 
Density functional theory (DFT) based calculations have proven  
highly successful in locating the muon implantation sites in several 
materials~\cite{moller2013,bonfa2013,bonfa2015,bonfa2016,sky2018}. 
Based on this, we performed structural relaxations by DFT to determine the muon stopping sites and, thus, get a better understanding of the evolution of $B_{\mu}$ with applied pressure and as a function of In/Sn substitution. Here, 
we consider the representative cases of $x=1.0$ and $x=0.5$.

To determine the muon implantation sites we used (0.45, 0.45, 0.54) 
as a starting position, since it corresponds to the {minimum of the electrostatic potential in \yps\ (see ~Apps.~\ref{app:D} and \ref{app:E}).
The final muon site resulting from structural relaxations, 
(0.459, 0.484, 0.50), is close to that obtained from the 
electrostatic potential minima, but has a higher site symmetry
($8j$ instead of $16l$).  Notably, we find 
4 symmetry-equivalent sites that are only 0.6~\AA\ apart from each other.  
At first approximation, we therefore assume the muon to occupy a delocalized region, spread over these equivalent closely-spaced 
sites. This implies the need to consider muon delocalization in 
the subsequent local dipolar-field calculations in the following form: $\langle {B} \rangle =\left<\phi_\mu \right| B \left| \phi_\mu \right> \approx \frac{1}{4} \sum_{i=1}^4 {B}_{i} $.
This procedure yields similar results to those where the muon is 
located at the centroid (0.5, 0.5, 0.5) position 
(of the four equivalent sites). To simplify the description, we report the results for this second case.

It is worth noting that, despite the presence of two transverse-field components in the $x = 0.3$ and 0.8 samples (suggesting 
two non equivalent muon sites), DFT calculations predict only one muon implantation site. Such apparent inconsistency could be 
resolved by considering the \emph{disorder} induced by either the In/Sn substitution or by applied pressure. 
Thus, at low temperatures, the whole sample could be pictured as an ensemble of nanometric size domains with either long- or short-range magnetic order. Such effect becomes increasingly relevant with increasing pressure, as suggested by the considerable rise in the local field width (see Fig.~\ref{fig:Delta_Bmu}). 
Magnetic frustration, expected to increase with increasing pressure, 
may further reinforce such effects. 
In this context, in the limit of even higher pressures, \ypis\ could 
be at the verge of a transition towards a nonmagnetic phase of a so-called ``valence bond solid'' (VBS), as already proposed in Ref.~\cite{Lacroix2011} 
to explain the vanishing of the AF phase above 40\,kbar in the $x = 1.0$ case 
\cite{Muramatsu2011}.

We evaluate the dipolar field at the muon site (due 
to Yb magnetic moments) by making the following assumptions: 
\textit{i}) $B_\mu$ is predominantly of dipolar origin.
\textit{ii}) the applied hydrostatic pressure, when lower than $p_{cr}$~(see App.~\ref{app:A}), distorts negligibly 
the atomic surrounding in the crystallographic unit cell. Thus, neither the muon implantation site(s), nor the propagation vector change significantly with pressure. 
\textit{iii}) The ordered Yb magnetic moment is 
set to 1\,$\mu_\mathrm{B}$, independently of the tin content and the applied pressure. This strong assumption is reasonably verified at ambient pressure,  where the Yb magnetic moment does not vary by more than 40\% (as deduced from neutron scattering measurements \cite{Bauer2010} at different In/Sn substitutions). 
Very recent neutron powder diffraction measurements as a function 
of pressure (up to 14\,kbar) on an $x = 0.6$ sample also show only an 
increase of 11\% of the ordered magnetic moment (App.~\ref{app:C}). 
Since the dipolar field contribution at the muon implantation site is linear with the ordered magnetic moment, the above 
tiny variations cannot justify the tenfold increase of $B_{\mu}$ reported 
in Fig.~\ref{fig:Bmu_Vmag}. Therefore, we can reasonably assume 
that the Yb magnetic moment is almost constant, at least up to $p_{cr}$.

Given the propagation vector $k=(0, 0, \frac{1}{2})$~\cite{Bauer2010, Mauro2018},  
the MAXMAGN code on the Bilbao crystallographic 
server~\cite{bilbcrys2014,PerezMato2015} identifies ten maximal magnetic space groups for the parent space group $P4/mbm$ (No.\ 127), allowing a nonzero field 
on Yb. These are labeled I to X and reported in Tab.~\ref{tab:table1} 
together with the corresponding magnetic structures on the Yb atoms.\\
For all these magnetic structures the dipole sum was calculated 
using a $100 \times 100 \times 100$ supercell.
In a perfect stoichiometric crystal, all the structures except
IV, VIII, and X have zero dipolar fields by symmetry.
However, Sn/In substitutions induce small in-plane displacements of the Yb atoms and of the muon position. The extent of 
these displacements was reproduced
through several random realizations of the $x=0.5$ concentration. 
The small dipolar field that arises as a consequence of the  
Yb displacements was evaluated by averaging on a set of 32 interstitial 
positions. This leads to local fields ranging from 3 to 10\,mT in  
those structures where the field would otherwise have been zero by symmetry.
The full set of results for the ten magnetic structures, obtained 
after averaging on the $x=0.5$ lattice realizations, is summarized 
in Tab.~\ref{tab:table1}. The additional averaging over the four 
equivalent muon sites has a negligible effect. Yet, for completeness, we report it in App.~\ref{app:E}.\\
Notably, the calculated dipolar field for the magnetic structure suggested by neutron scattering at ambient pressure [structure VI, 
observed for $x(\mathrm{Sn})=0.5$--0.9] \cite{Bauer2010,Mauro2018} 
is found to be about 7\,mT. This field value agrees well with those 
sensed by muons in the Sn/In substituted samples at ambient-pressure, 
hence justifying our original omission of the contact-field contribution. 
At the same time, the above result suggests  
that the Yb magnetic moments cannot remain in the magnetic structure VI also at high pressures, where 
the measured internal field is four-to-ten times the expected one. For the same reason, none of the magnetic structures I--III, V, VII, and IX can describe the Yb ordering observed at high pressure for 
$x=0.3$, 0.6, and 0.8, since in all these cases the calculated dipolar fields range from 3 to 7\,mT.\\
Out of the ten long-range orders considered here, only three 
justify the drastic increase of local field at the muon site for $p>p_\mathrm{cr}$. These include the magnetic structures IV, VIII, and X, marked in bold in Tab.~\ref{tab:table1} and shown in Fig.~\ref{fig:magstruc2}.
However, since also a contact hyperfine component may be present, we refrain from making direct comparisons between the measured and calculated local field.\\
To summarize, our results strongly suggests a \emph{reordering of the Yb magnetic moments} at high pressures, above $p_\mathrm{cr}$. The above-mentioned recent low-temperature neutron and room temperature synchrotron  
diffraction data (see {Ref.~\cite{Mauro2018} and Apps.~\ref{app:A} and \ref{app:C}), both collected under applied pressure, 
suggest \textit{no changes in the magnetic and crystalline structure up to} $p_\mathrm{cr}$. However, above $p_\mathrm{cr}$ a structural transition does take place. 
Consequently, we considered also the supposed high-pressure 
monoclinic structure $P2_1/c$ (App.~\ref{app:D}). 
In this case, four irreducible representations allow for a non-zero magnetic moment on the Yb ion. 
If the Yb moments are constrained to lie
either along $a$ or in the $bc$ plane, with $|m_b|=|m_c|$,
the local field at the muon site does not change appreciably with 
respect to the values identified for the tetragonal low-pressure 
structure. This implies that the structural transition at $p_\mathrm{cr}$ does not appreciably change the calculations of the internal field $B_{\mu}$ done for the ten models deduced for the tetragonal symmetry summarized in Tab.~\ref{tab:table1}.  

At the same time neutron powder diffraction in the 
$x = 0.6$ case revealed a stable magnetic structure  
up to 14\,kbar, suggesting that a pressure-induced magnetic phase transition --- if any --- could take place only at very high pressures, beyond the instrument limits. At intermediate pressure values, we expect the effects of the different competing AF-ordered phases,  
to correspond to the magnetic structures IV, VIII, and X mentioned above. These competing orders  could result in magnetic frustration and, hence, significantly enhance the local field width at high pressure. At the same time, all these magnetic structures are compatible with the large increase of magnetic field observed above $p_\mathrm{cr}$ [see Fig.~\ref{fig:3D_diag}(b)]. 

\begin{figure}[tbh]
	\centering \textbf{(a) Struc. IV }   {\hspace{5em}}     \textbf{(b) Struc. VI}\\
	\includegraphics[width=0.195\textwidth]{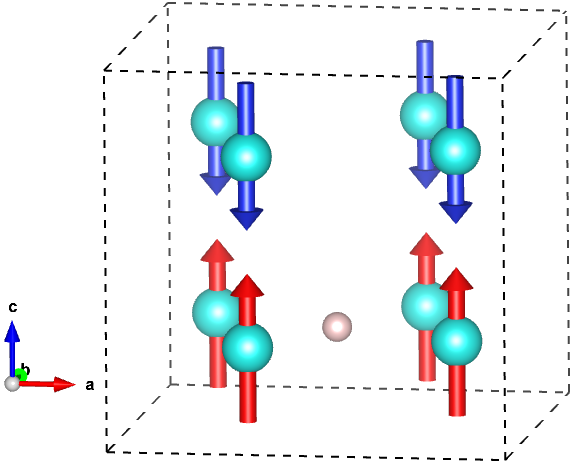} 
	\includegraphics[width=0.195\textwidth]{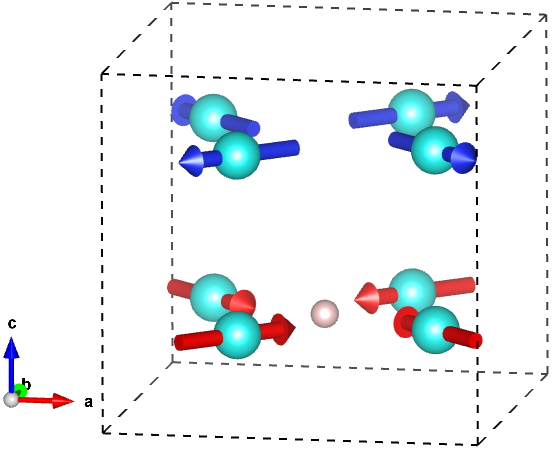}
	\\[\baselineskip]
	\centering \textbf{(c) Struc. VIII}     {\hspace{5em}}   \textbf{(d) Struc. X}\\
	\includegraphics[width=0.195\textwidth]{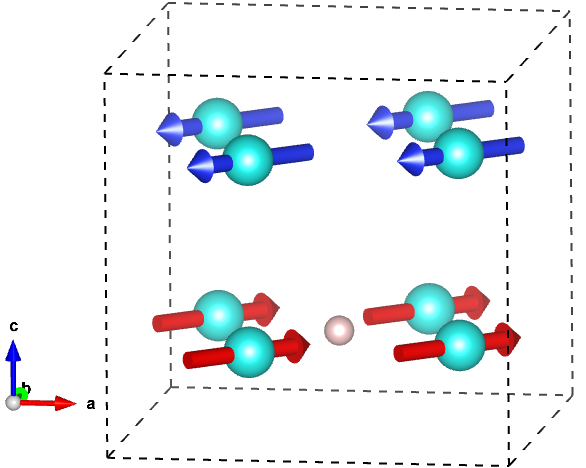}
	\includegraphics[width=0.195\textwidth]{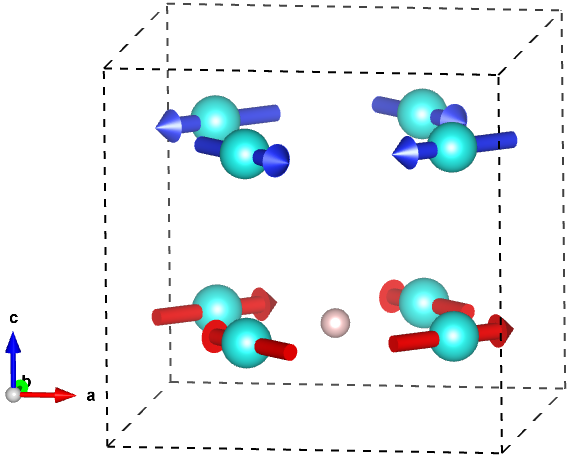}
	\caption{\label{fig:magstruc2}Magnetic structures IV (a), VI (b), 
		VIII (c), and X (d) listed in Table~\ref{tab:table1} for a doubled 
		unit cell along $c$. Structure VI is the known AF structure at 
		ambient pressure, while structures IV, VIII, and X justify the 
		local field $B_\mu$ we measure 
		at high pressure. For clarity only 
		the Yb atoms and the muon (pink sphere) are shown. 
		The structures were drawn using   
		VESTA~\cite{vesta2008}.}
\end{figure}

\section{Discussion} The ground-state electronic properties of heavy-fermion compounds 
are determined by a subtle interplay between the 
RKKY and Kondo interactions, whose characteristic temperature (energy) 
scales are $T_\mathrm{RKKY} \propto g^2$ and $T_\mathrm{K} \propto e^{-1/g}$, 
with $g =  \Gamma/ \pi \Delta_f$, a coupling parameter.
Here $\Gamma \propto V_{fc}^2$, where $V_{fc}$ represents the 
overlap between the wave functions of $4f$ and conduction electrons, while 
$\Delta_f$ is the excitation energy between the magnetic and non-magnetic 
electronic configurations \cite{Muramatsu2011}. 
In most Yb systems, valence fluctuations and the reduced size of $4f$ Yb ions  
rule out significant changes of $V_{f\!c}$ up to 25\,kbar \cite{Muramatsu2011}. 
Consequently, in the considered pressure range, 
the coupling strength $g$ is determined almost entirely  
by the excitation energy $\Delta_f$.

At the boundaries of the $p$-$T$-$x$ phase diagram, i.e., at low 
hydrostatic pressures in the pure In or Sn case (i.e., for $x \sim 0$ 
or 1), the excitation energy $\Delta_f$ is expected to assume its minimum 
value, which maximizes $g$ \cite{Muramatsu2011}. In turn, this is 
reflected in $T_\mathrm{K}$ values up to 20--30\,K, resulting in a 
non\-mag\-net\-ic ground state with non-Fermi-liquid  
properties, as confirmed by our \musr\ measurements 
(see Fig.~\ref{fig:3D_diag}). 
In this case, our Yb-based Kon\-do-lat\-ti\-ce system is found in the 
Kondo screening region of the Doniach phase diagram \cite{Doniach1977}.

The application of hydrostatic- or chemical pressure implies a reduction of the 
unit cell volume \cite{Bauer2005}. 
Moderate hydrostatic pressure has negligible effects on the Yb$^{3+}$ valence, 
as confirmed by x-ray absorption spectroscopy in the partial 
fluorescence yield mode (PFY-XAS) and by resonant x-ray emission 
spectroscopy (RXES) under applied pressure \cite{Yamaoka17}. Also 
the In/Sn substitution, (chemical pressure), does not distinctly affect 
valence, since dc susceptibility measurements indicate almost
constant Yb magnetic moments in the paramagnetic state, independent of tin concentration {(App.~\ref{app:B}). 
In this region of the phase diagram, an increase of pressure induces a
growth of resistivity \cite{Bauer2004,Bauer2005}, corresponding to a 
reduced $f$-$c$ electron coupling and to an increased excitation energy $\Delta_f$. 
Both of them stabilize the antiferromagnetic phase \cite{Muramatsu2011}, 
clearly suggesting that pressure (either hydrostatic or chemical) shifts 
a Kon\-do-lat\-ti\-ce system towards the RKKY interaction regime 
in the Doniach phase diagram \cite{Doniach1977}.

As shown in Fig.~\ref{fig:3D_diag}(a), the antiferromagnetic phase covers a wide area of the $p$-$T$-$x$ phase diagram. At ambient pressure it spans only the $0.3<x<0.9$ range. However, upon increasing pressure, the AF region extends to cover the full $0 < x < 1$ range, 
albeit with a slightly 
asymmetric distribution of $T_\mathrm{N}$ temperatures. 
Thus, for $x=0$, the system is at the verge of a magnetic transition 
at a pressure of 25\,kbar, while for $x=1$ the AF phase has its onset at only about 10\,kbar.
The fine balance between chemical pressure and effective doping, 
here realized through the In/Sn substitution, may explain 
the slightly off-centered dome shape in the $p$-$T$-$x$ phase diagram.  In addition, M\"ossbauer studies evidence different types of ground states in the pure In and Sn systems \cite{Muramatsu2011,ActaPol2003}. Off-centered features might thus result naturally. 

Similar features are also found in the $p$-$B_{\mu}$-$x$ phase 
diagram, shown in Fig.~\ref{fig:3D_diag}(b).
Interestingly, for $x<1$, a marked change in the slope of the $B_\mu$ vs.\ $p$ plot 
is clearly  visible at about $p_\mathrm{cr}$ (see continuous lines). Such  
pressure value coincides with that where a structural phase transition 
was shown to occur at ambient temperature (see Fig.~\ref{fig:SM_1}) for any $x(\mathrm{Sn})$ 
value, except for $x=1$. In this case, no structural transitions 
could be detected (up to 100\,kbar). In this case, we note also 
that the internal field $B_\mu$ never exceeds the values found at 
ambient pressure in the AF dome~\cite{MauroXray}.
Consequently, the significant increase of local magnetic field observed in 
the central part of the $p$-$B_{\mu}$-$x$ phase diagram cannot be 
ascribed to a distinct change of the Yb magnetic moment, the latter being 
ruled out by our dipolar-field calculations.   
We suggest, instead, that the steep increase in the local magnetic field 
is due to a \emph{reorientation} of the Yb magnetic substructure and/or to magnetic frustration resulting from the \emph{competition 
of at least three different AF magnetic structures.} 
Since such reordering occurs at $p > p_\mathrm{cr}$, it is very 
tempting to assume that the change in magnetic structure  
is most likely driven by the above mentioned structural transition. 

Finally, we would like to address the significant increase in $T_\mathrm{N}$ with increasing pressure. To this aim it is 
instructive to compare the quantum critical behavior of Ce- with Yb-based 
systems \cite{Flouquet2012,Braithwaite2013}. Both Ce and Yb exhibit a 
trivalent magnetic state (Ce$^{3+}$, 4$f^{1}$ vs.\ Yb$^{3+}$, 4$f^{13}$), 
which can fluctuate to a nonmagnetic state, corresponding to an empty 
$4f^{0}$ shell for Ce and to a filled ($4f^{14}$) shell for Yb. 
To a first approximation, in both cases pressure leads to a 
delocalization of the $4f$ electrons. However, while this drives Ce towards a 
non\-mag\-net\-ic 4$f^{0}$ state, in Yb it favors the magnetic 
4$f^{13}$ state, in an almost specular behavior, known as the mirror-like behaviour of cerium and ytterbium. This implies a reinforced magnetic order in Yb-based systems under 
applied pressure, as indeed observed, e.g., in YbCu$_{2}$Si$_{2}$ 
\cite{Panella2011} or in YbRh$_{2}$Si$_{2}$ \cite{Knebel2006}, and 
to its suppression in Ce-based systems as, e.g., in CeRh$_{2}$Si$_{2}$ 
\cite{Knebel2006} (where an initial $T_\mathrm{N}$ of 35\,K goes to 
zero in only 12\,kbar).

\section{CONCLUSIONS} The intriguing magnetic behavior  
of \ypis\ was studied mostly via \musr\ spectroscopy and elucidated 
by detailed DFT calculations. Both hydrostatic- and chemical 
pressure (the latter through In/Sn substitution) promote 
an AF coupling of Yb$^{3+}$ magnetic moments. 
The internal field evolution with pressure suggest a possible reorientation of the Yb moments and/or the presence of frustrated magnetism, most likely due to different competing AF-interactions above $p_\mathrm{cr}$ and 
is presumably driven by a structural phase transition.\\

\section{ACKNOWLEDGMENTS}
GL acknowledges financial support from the CNR Short-Term 
Mobility Program for his stay at the University of Pre\v{s}ov (Slovakia), 
where part of the magnetization measurements were performed. 
Part of this work was supported by the Schwei\-ze\-rische Na\-ti\-o\-nal\-fonds 
zur F\"{o}r\-de\-rung der Wis\-sen\-schaft\-lich\-en For\-schung (SNF) under 
grant No.\ 200021-169455. RDR  acknowledges support from the European 
Union's Horizon-2020 research and innovation program (grant No.\ 654000). 
RDR, IJO, and PB also acknowledge the computing resources provided 
by the Swiss National Super\-computing Centre (CSCS) (project sm16), 
CINECA (project IsC58), STFC SCARF cluster, UK, and the HPC at the University of Parma, Italy. MR, IC, AD, and GP were supported by the grants VEGA  1/0956/17, VEGA 1/0611/18, and APVV-16-0079. The authors  acknowledge Elettra for allocation of HP-XRPD beamtime and ILL for beam time allocation under the experiment code 5-31-2584. Neutron Diffraction data are available from ILL at DOI: 518 10.5291/ILL-DATA.5-31-2584.


\appendix
\section{X-ray diffraction}\label{app:A}

The synchrotron x-ray powder diffraction (XRPD) measurements were 
carried out at 300\,K at the Xpress beamline of Elettra (Trieste, Italy). Two datasets were collected: at ambient pressure using a radiation wavelength 
$\lambda = 0.700$\,\AA\ (for $x = 0.0$, 0.3, 0.4, 0.6, 0.8), and as a function of pressure --- up to $\sim 35$\, kbar --- using $\lambda = 0.496$\,\AA\ (for $x = 0.0$, 0.3, 0.8). 
The ambient-pressure XRPD data could be satisfactorily fitted using a tetragonal $P4/mbm$ space group across the whole compositional range, in agreement with previous investigations \cite{Bauer2005,Mauro2018}.

The XRPD patterns collected under applied pressure at the Xpress beamline reveal that all the 
investigated samples undergo a structural transition as the pressure increases (Fig.~\ref{fig:SM_1}). Yb$_2$\-Pd$_{2}$In ($x=0$) exhibits a sharp structural transition at $\sim 6.5$\,kbar. In the $x=0.3$ case, the structural transition has its onset at about the same pressure, but develops over a wider pressure range, to complete at $\sim 15.0$\,kbar, thus indicating the first-order character of the transition. For $x=0.8$, the structural transition is triggered at a significantly higher pressure ($\sim 16$\,kbar). By inspecting the above data one can 
deduce that, to some extent, the Sn substitution hinders the formation of the high pressure (HP) phase.

\begin{figure}[tbh]
	\centering
	\includegraphics[width=0.45\textwidth]{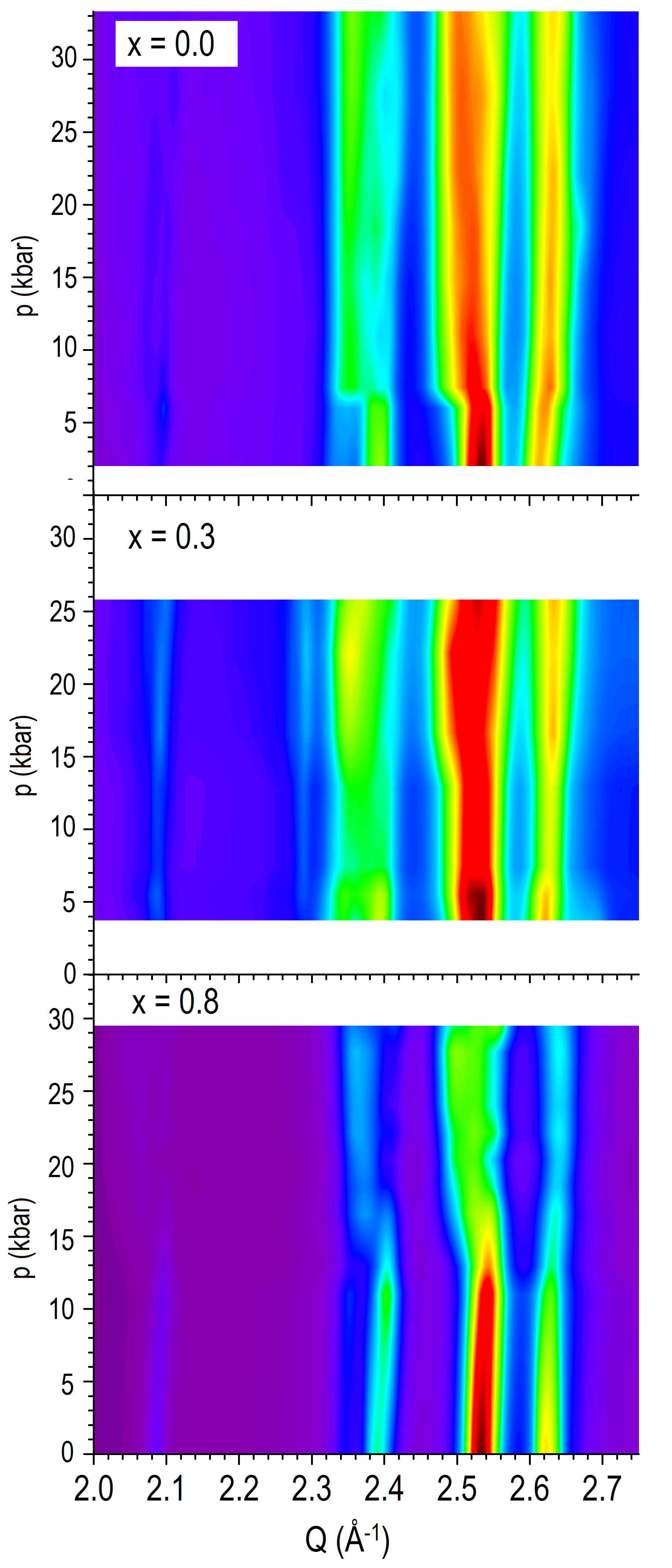}
	\caption{\label{fig:SM_1}XRPD patterns as a function of pressure for the $x=0$, 0.3, and 0.8  samples. Here we show an enlarged view of the $Q$-region where the highest intensity peaks are located and their evolution with the applied pressure.}
\end{figure}
Attempts to ascertain the crystal structure of the HP phase were carried out, notwithstanding the unavoidable background due to
the experimental set-up. To this purpose, the HP-XRPD data of the $x=0$ case were carefully analysed, since it exhibits not only a complete structural transformation at lower pressures, but also because of the lack of a diffraction line broadening due to In-substitution.
Figure~\ref{fig:SM_2} shows the evolution with pressure of the peak located at 
$Q \sim 2.09$\,\AA$^{-1}$, representing the 111 reflection of the low pressure (LP) phase that splits in the HP polymorph. Such behavior provides some clues and allows us to exclude some of the possible structural models. Firstly, the peak splitting indicates the suppression of the 4-fold rotational symmetry. Among the maximal subgroups of the $P4/mbm$ space group, only the orthorhombic space groups $Cmmm$ and $Pbam$ do not contain a 4-fold rotational axis. Nevertheless, none of them can account for the observed splitting. Upon further inspection of the structural models pertaining to the $Cmmm$ and $Pbam$ subgroups, it turns out that the HP phase must be at least monoclinic to reproduce such peak splitting. 
\begin{figure}[tbh]
	\centering
	\includegraphics[width=0.45\textwidth]{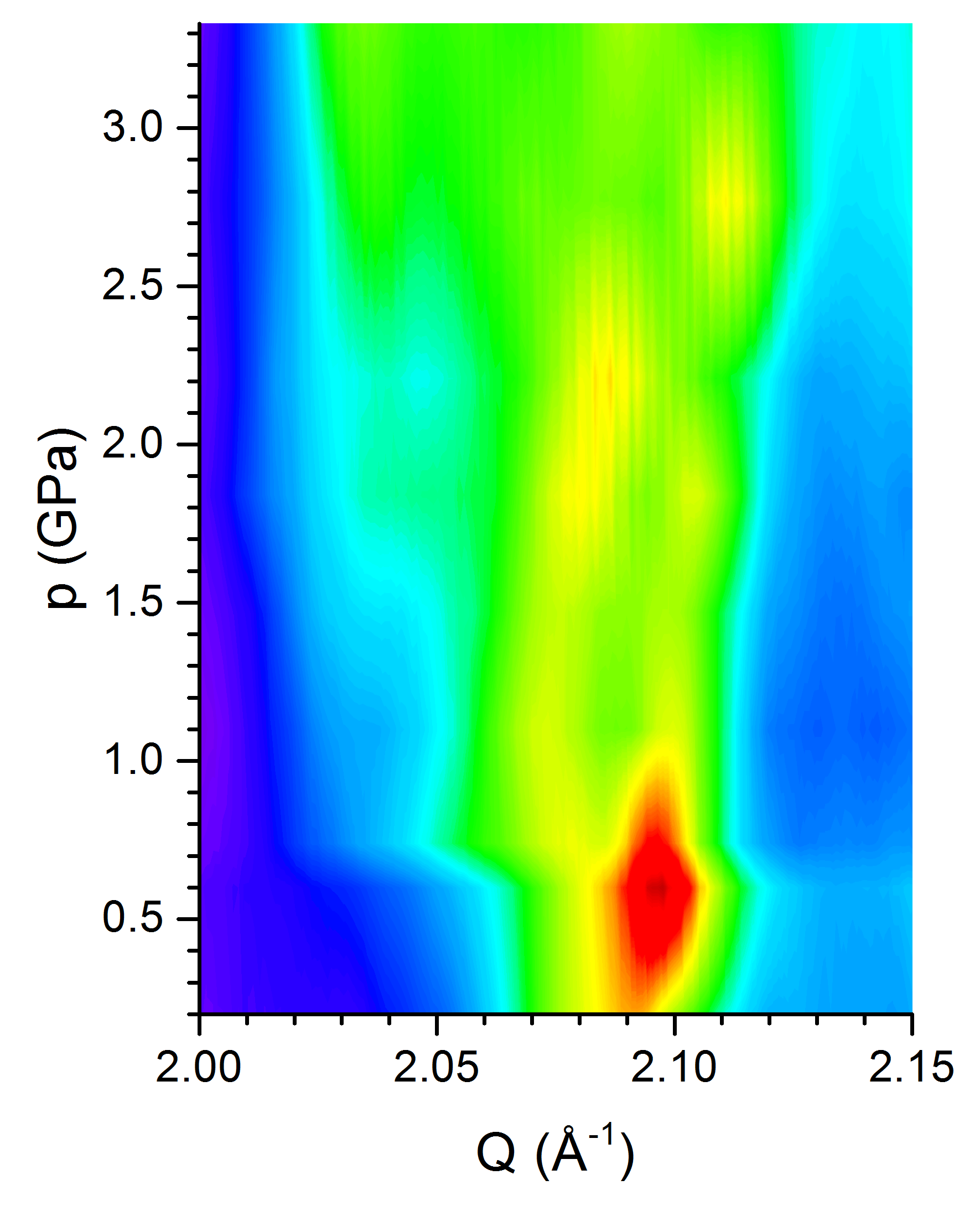}
	\caption{\label{fig:SM_2}XRPD data showing the evolution with pressure of the diffraction peak located at $Q ~ 2.09 \AA^{-1}$ for the $x=0$ case. The 111 reflection pertaining to the low-pressure polymorph splits at a higher pressure, $p_{cr}$, which marks the structural transition.}
\end{figure}
Several monoclinic models were tested and the best fit of the experimental 
data was obtained with a $P2_1/c$ structural model. The corresponding Rietveld refinement plot and structural data are reported in Fig.~\ref{fig:SM_3} 
and Table \ref{tab:S_1}, respectively. Noteworthy, the $R$-values listed in Table \ref{tab:S_1} are somewhat biased towards relatively low values due to the strong instrumental contribution to the diffraction pattern. Therefore, the proposed structural model could differ significantly from the real structure. 
\begin{figure}[tbh]
	\centering
	\includegraphics[width=0.49\textwidth]{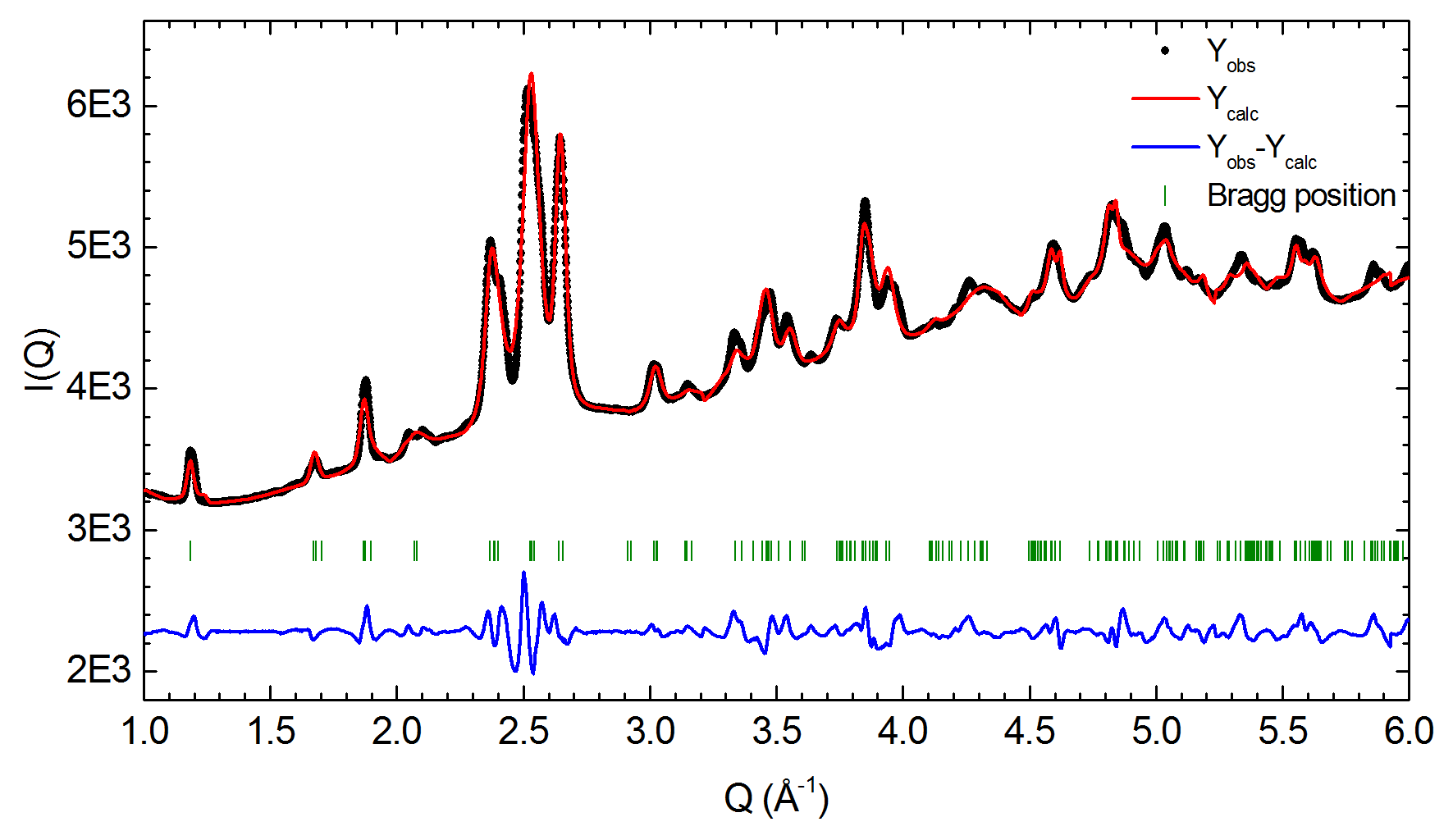}
	\caption{\label{fig:SM_3}Rietveld refinement plot for Yb$_2$Pd$_2$In 
		using XRPD data collected at 33.3\,kbar and the monoclinic $P2_1/c$ space group.}
\end{figure}
\begin{table}[tbh]
	\centering
	\renewcommand{\arraystretch}{1.2}
	\caption{\label{tab:S_1}Yb$_2$Pd$_2$In structural parameters, as 
		refined from XRPD data collected at 290\,K and 33.3\,kbar (space group 
		$P2_1/c$).}
	\begin{ruledtabular}
		\begin{tabular}{p{1mm}lccccp{1mm}}
			\vspace{2mm}
			& \lower 0.35mm \hbox{$a$\,[\AA]}
			& \lower 0.35mm \hbox{$b$\,[\AA]}
			& \lower 0.35mm \hbox{$c$\,[\AA]}
			& \lower 0.35mm \hbox{$\beta$\,[deg.]}\\[2pt]
			& \lower 0.35mm \hbox{3.6874(1)}
			& \lower 0.35mm \hbox{7.5276(1)}
			& \lower 0.35mm \hbox{7.4740(1)}
			& \lower 0.35mm \hbox{89.65(1)}\\[5pt]
			\bottomrule 
			& \lower 0.35mm \hbox{Atom}
			& \lower 0.35mm \hbox{Wyckoff site}
			& \lower 0.35mm \hbox{$x$}
			& \lower 0.35mm \hbox{$y$}
			& \lower 0.35mm \hbox{$z$}\\[2pt]
			\midrule 
			&Yb &   4$e$ & 0.5334(3) & 0.6679(1) & 0.8210(1)\\
			&Pd &   4$e$ & 0.0975(2) & 0.1334(1) & 0.3787(1)\\
			&In &   2$a$ & 0         & 0         & 0        \\
			\bottomrule 
			& \multicolumn{5}{c}{$R_\mathrm{Bragg} = 2.40$, \quad $R_\mathrm{factor} = 1.59$} & \\
			
		\end{tabular}
	\end{ruledtabular}
\end{table}

\begin{figure}[tbh]
	\centering
	\includegraphics[width=0.55\textwidth]{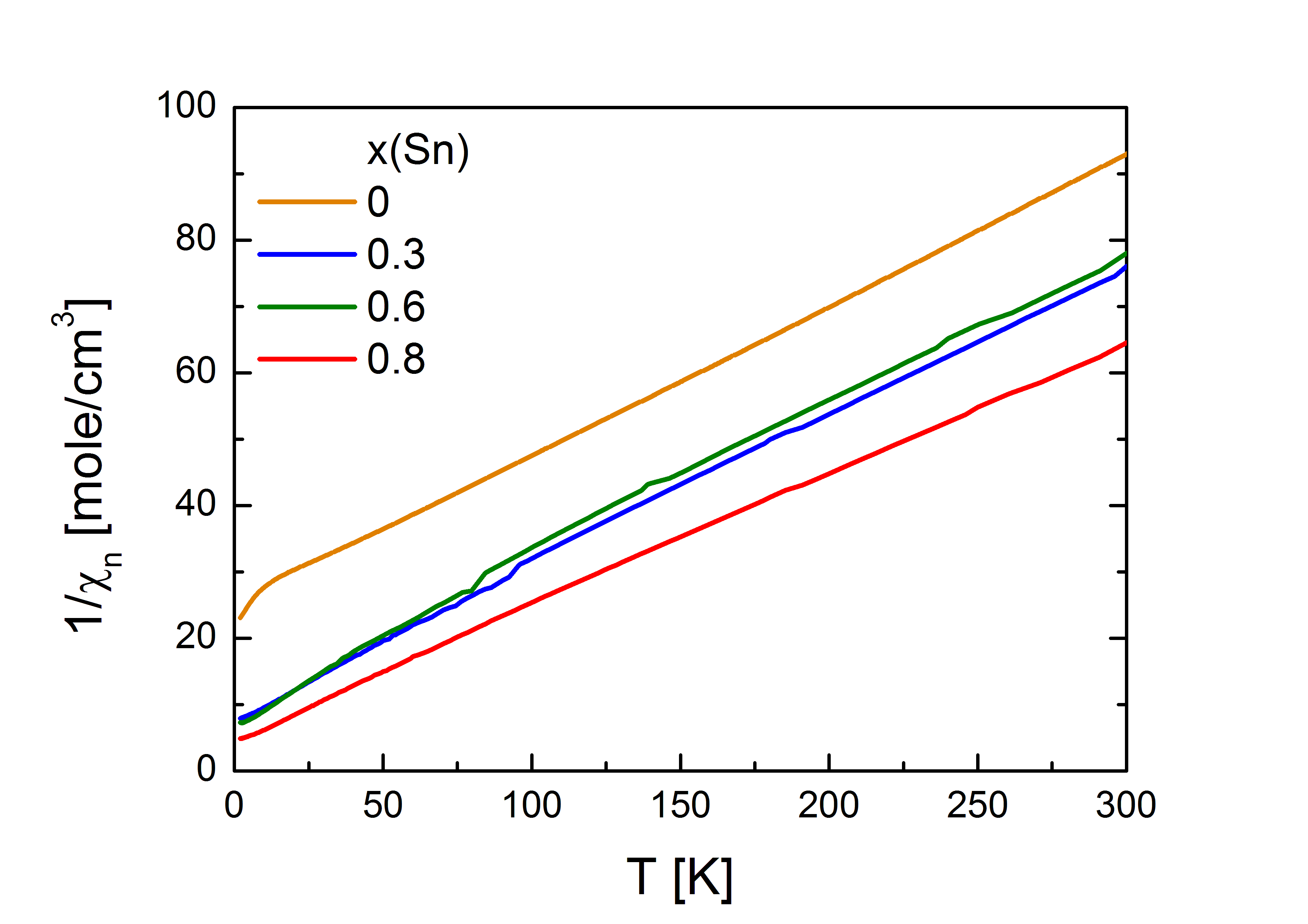}
	\caption{\label{fig:SM_5}Temperature dependent inverse molar susceptibility for the investigated samples. The almost straight lines in indicate a Curie-Weiss-like behaviour (see text).}
\end{figure}

\section{DC magnetization}\label{app:B}
To assess the bulk magnetic properties of the as-grown samples, 
systematic dc susceptibility measurements were performed by means of an MPMS and a VSM Dynacool magnetometer systems, both from Quantum Design. $1/\chi(T)$ curves measured from 2 to 300\,K at $\mu_0 H = 3$\,T are reported in Fig.~\ref{fig:SM_5}.
Data exhibit a clear linear behavior down to about 30--50\,K, below which they show a downward curvature. 
Most likely, this reflects the increasingly stronger correlations 
between Yb  magnetic moments as one approaches the AF ordering temperature. 
After excluding the low-$T$ data, the magnetic susceptibility 
between 50 and 300\,K could be fitted by means of a modified Curie Weiss law: 
\begin{equation}
\chi_\mathrm{mol}(T)=\frac{C}{T-\theta}+\chi_{0}.  
\end{equation} 
Here, $C$ is the Curie constant, $\theta$ is the Curie-Weiss temperature, and $\chi_0$ is a constant that accounts for the different temperature-independent contributions. 
In our case,  $\chi_0$ is dominated by the Pauli paramagnetism of the conduction electrons. 

Figure~\ref{fig:SM_6} summarizes the fit parameters as a function of Sn content $x$ and highlights some interesting features. Firstly, the effective magnetic moment is close to 4.53\,$\mu_\mathrm{B}$, the expected value for $4f^{13}$ ($^{2}F_{7/2}$-term) magnetic Yb$^{3+}$ ions, and is mostly independent of the In-Sn substitution rate. Secondly, the Curie 
\begin{figure}[tbh]
	\centering
	\includegraphics[width=0.4\textwidth]{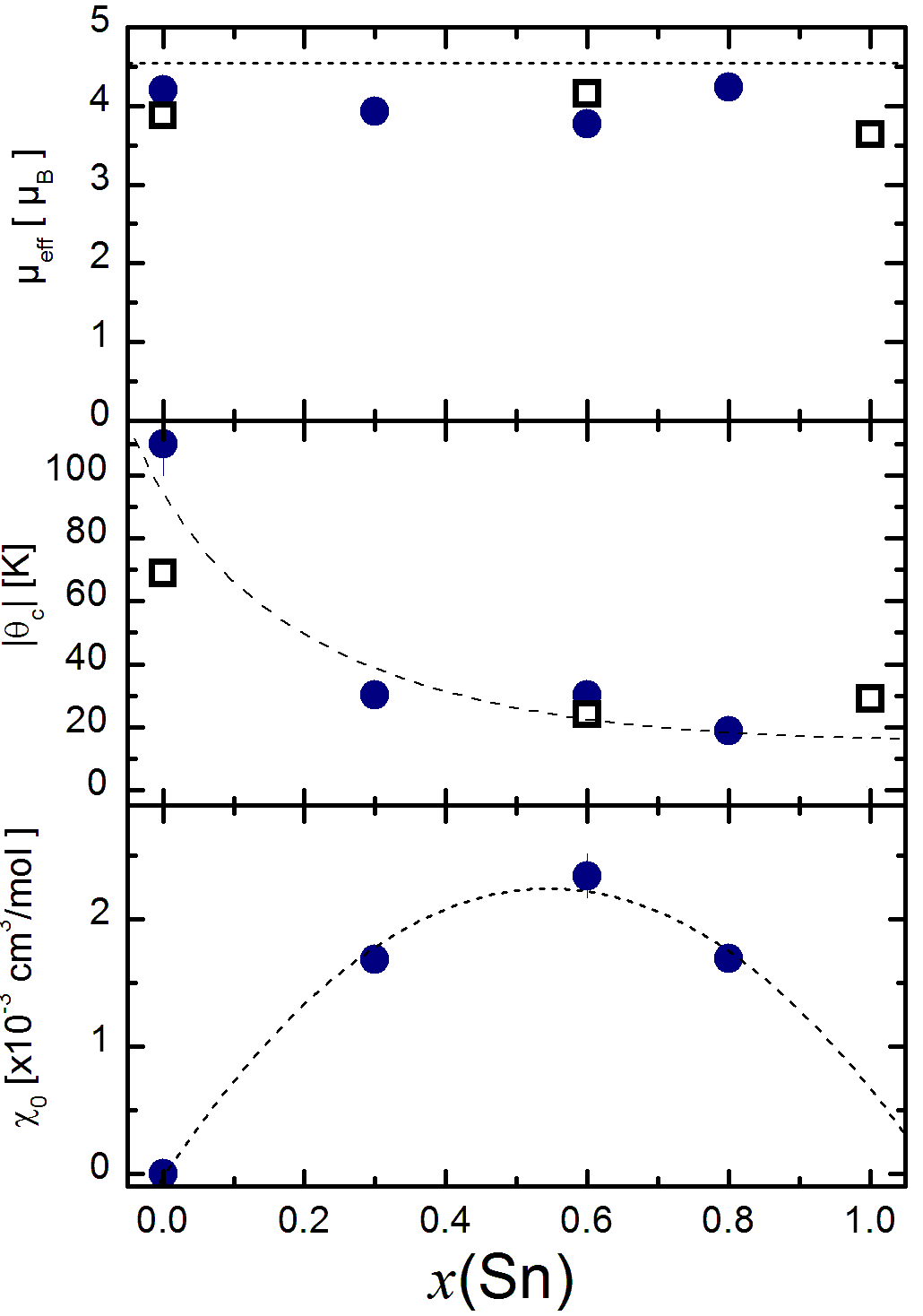} 
	\llap{\parbox[b]{20mm}{\large{(a)}\\\rule{0ex}{69mm}}}
	\llap{\parbox[b]{22mm}{\large{(b)}\\\rule{0ex}{43mm}}}
	\llap{\parbox[b]{22mm}{\large{(c)}\\\rule{0ex}{13mm}}}
	\caption{\label{fig:SM_6}Fit parameters, as extracted from Curie-Weiss 
		fits of the dc magnetic susceptibility data, vs.\ Sn content $x$. Top panel: effective Yb magnetic moment. The dashed line represents the expected 
		magnetic moment per Yb$^{3+}$ ion. Middle panel: Curie-Weiss temperature. The dashed line is a guide for the eyes. Bottom panel: $T$-independent contribution $\chi_0$ and parabolic fit. In the first two cases, blue filled circles refer to data from this study; black empty squares are data from Ref.~\onlinecite{Bauer2005}.}
\end{figure}
temperature is always negative, thus suggesting a predominance of \emph{antiferromagnetic-like interactions} between Yb$^{3+}$ ions. Finally, the $T$-independent term $\chi_0$ shows a parabolic dependence vs.\ Sn content. In particular, it exhibits magnitudes in the $10^{-3}$\,cm$^3$/mol range, with the maximum being reached at optimum doping. We recall that for simple metals $\chi_0$ is about $10^{-6}$\,cm$^3$/mol \cite{Ashcroft1976},  yet it may increase by more than a factor of 1000 in heavy-fermion (HF) compounds \cite{Gegenwart2005}. This is also our case, where the Yb $4f$ electrons hybridize with the conduction electrons, producing a strong enhancement of the carriers' effective mass (heavy-fermion state).

\section{Neutron powder diffraction}\label{app:C}
The magnetic structure of \ypis\ at ambient pressure, as resulting from neutron diffraction data \cite{Mauro2018} on an optimally-doped compound, is shown in Fig.~\ref{fig:magn_struct}. The anticollinear in-plane ordering of the Yb${}^{3+}$ ions, as well as 
their an\-ti\-fe\-rro\-mag\-net\-ic intra\-lay\-er interactions 
are in agreement with the magnetometry results.
\begin{figure}[tbh]
	\centering
	\includegraphics[width=0.5\textwidth]{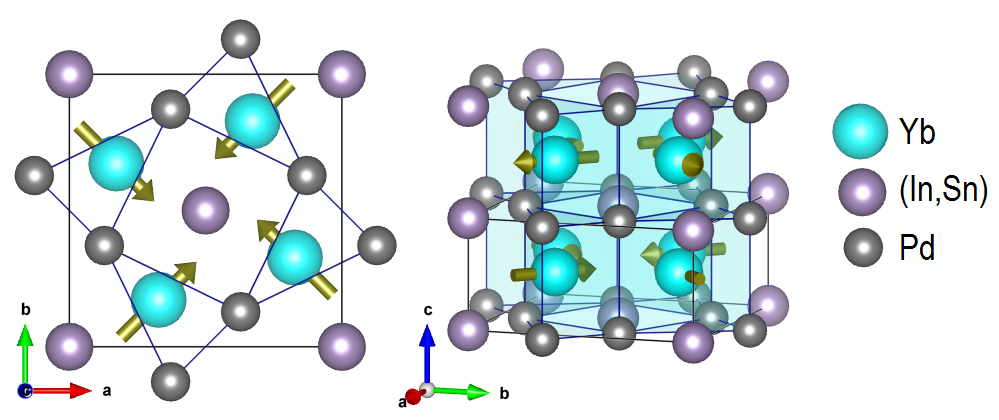} 
	\llap{\parbox[b]{30mm}{\large{(a)}\\\rule{0ex}{0mm}}}
	\rlap{\parbox[b]{40mm}{\large{(b)}\\\rule{0ex}{0mm}}}
	\caption{\label{fig:magn_struct}Magnetic structure of \ypis\ at ambient pressure, as derived from neutron scattering 
		data \cite{Mauro2018}. Note the anticollinear 
		in-plane order (a) and the antiferromagnetic intra-layer arrangement (b) of the 
		magnetic Yb${}^{3+}$ (4$f^{13}$) ions.}
\end{figure}
Because \musr\ measurements reveal that in all the cases (and in particular for $x = 0.6$)  the magnetic phase is greatly enhanced by external pressure, we selected the $x=0.6$ sample for further neutron powder diffraction (NPD) analysis under applied pressure. To this aim, 
NPD measurements were performed at the D20 diffractometer of the Institute Laue-Langevin (Grenoble, France). A TiZr clamp cell was loaded with about 4\,g of sample, using Fluorinert as a pressure transmitting medium. Data were collected at 300\,K, as well as in the 1.5--5.0\,K range (above and below the magnetic transition) using an orange cryostat ($\lambda = 2.4174$\,\AA). To determine the pressure, a small amount of Pb was added to the sample. The applied pressure was thus directly calculated by substituting the measured structural parameters of Pb into its equation of state.
\begin{figure*}[tbh]
	\centering
	\large{(a)}
	\includegraphics[width=0.47\textwidth]{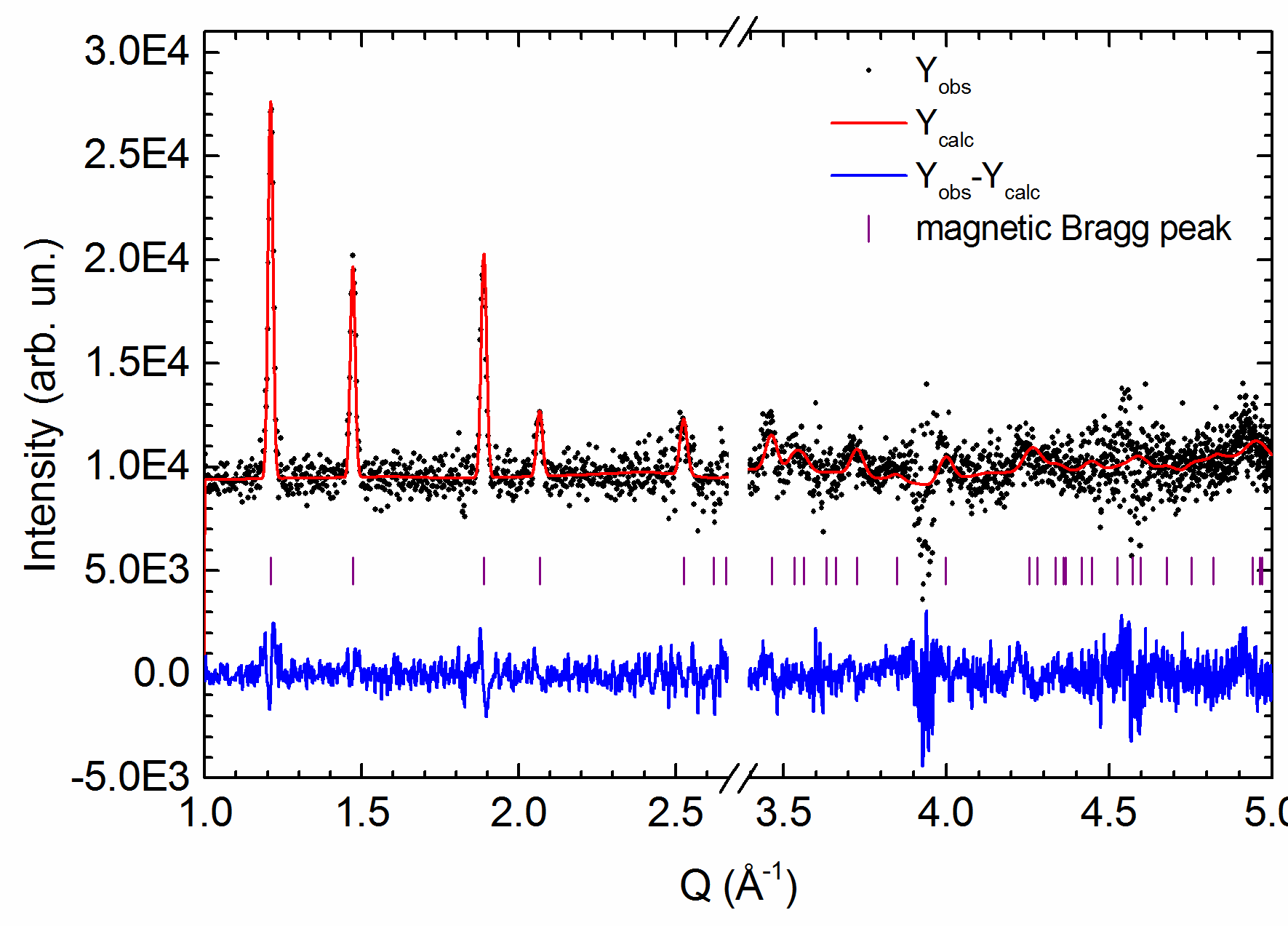} 
	\large{(b)}
	\includegraphics[width=0.45\textwidth]{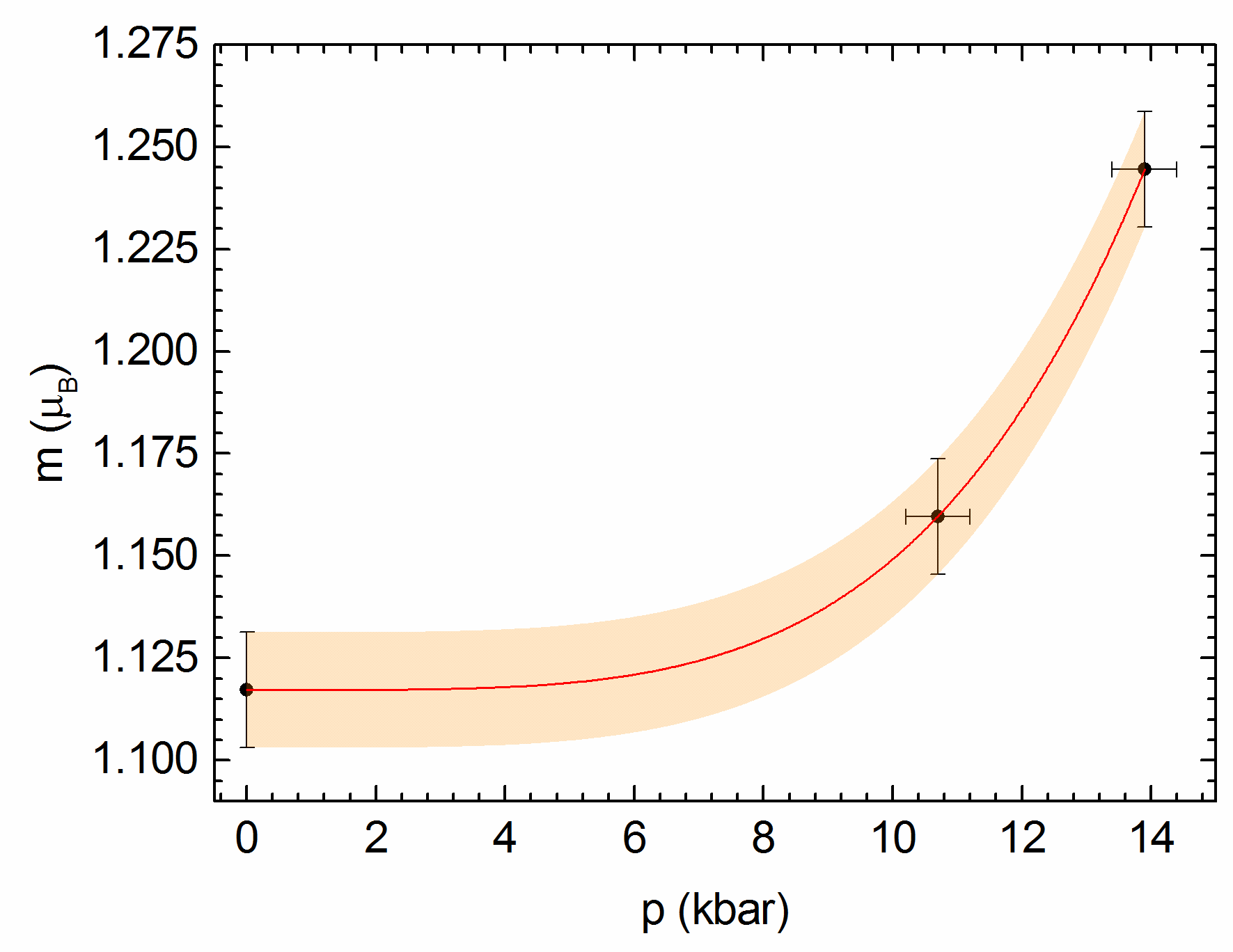} 
	\caption{\label{fig:HP-NPD}(a) Coherent magnetic scattering 
		in an $x = 0.6$ sample, measured at 13.9\,kbar and 1.5\,K, and its Rietveld 
		refinement (see text for details). (b) Evolution with pressure of the of Yb$^{3+}$ magnetic moment as derived from Rietveld fits.}
\end{figure*}
Firstly, it is worth noting that the low resolution setup used to collect the NPD data prevented us from detecting the structural transformation taking place at high pressure. The Rietveld refinement was carried out by fitting the difference between the NPD data collected at 10\,K (in the paramagnetic state) and at 1.5\,K (in the magnetically ordered state), as shown in Fig.~\ref{fig:HP-NPD}(a). No clear evidence of a magnetic phase transition with increasing pressure was observed up to $\simeq 14$\,kbar, a value where the magnetic structure adopted at  ambient pressure \cite{Mauro2018} is still retained. Nonetheless, we still could detect a moderate increase of the ordered magnetic moment with pressure, 
as derived by the Rietveld fits shown in Fig.~\ref{fig:HP-NPD}(b).
\section{Details of the DFT calculations}\label{app:D}
\begin{figure}[tbh]
	\centering
	\large\textbf{(a)}
	\includegraphics[width=0.34\textwidth]{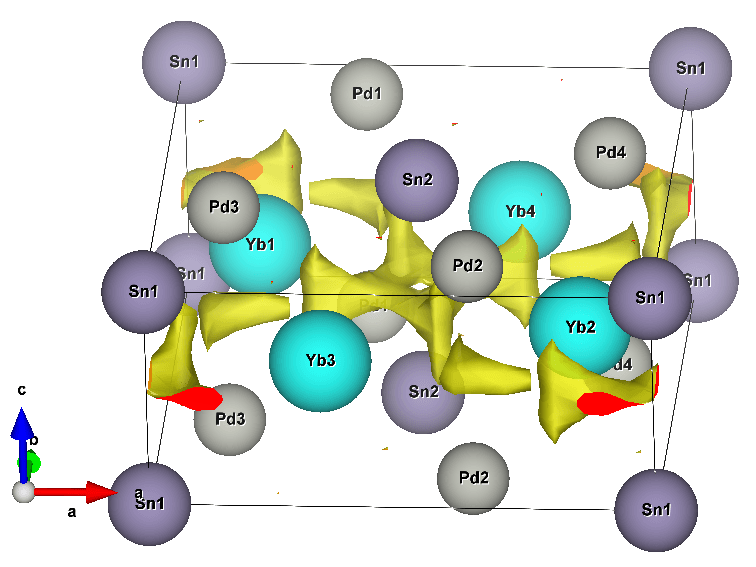}
	\vspace{5mm}\\
	\large\textbf{(b)}
	\includegraphics[width=0.34\textwidth]{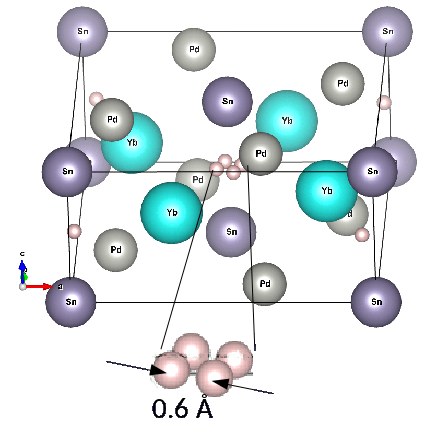} 
	\caption{\label{fig:musite1}Muon-stopping sites as obtained from the minima of the electrostatic potential (a) and from 
		self-consistent DFT calculations (b). In the latter case the perturbation caused by  muons to the lattice is also accounted for.}
\end{figure}
To determine the muon site(s) in \ypis{}, we adopted the following strategy: firstly, we calculated the electrostatic potential in the unit cell. The resulting minima were then used as starting trial positions for the muon, treated in this second step as an impurity in a supercell where all atomic positions are allowed to relax.

The electrostatic potential was calculated using the non-magnetic \yps\ unit cell, 
which belongs to the $P4/mbm$ space group, with lattice parameters $a = b = 7.5789$\,\AA\ and $c = 3.6350$\,\AA. Here, Yb ions occupy the $4h$ Wyckoff 
positions at (0.1724, 0.6724, 0.5), Pd the $4g$ positions at 
(0.3716, 0.8716, 0.0) and Sn the $2a$ positions at (0.0, 0.0, 0.0). The potential minima are shown in the isosurface plot of Fig.~\ref{fig:musite1}(a). 
In a second step, the muon was introduced as a hydrogen impurity at the potential minimum, corresponding to the interstitial position (0.45, 0.45, 0.54). 

A non-magnetic 2$\times$2$\times$4 supercell containing 160 atoms was used to model both \yps\ and \ypis . Random distributions of In/Sn atoms [$x=0.5$] were used to account for the substitution. 
The plane\-wa\-ve- and pseudopotential based implementation provided by the Quantum Espresso suite of codes \cite{qe2009} was used for the structural relaxation.
For the exchange correlation functional, the Perdew-Burke-Ernzerhof (PBE)  \cite{pbe1996} parametrization was used.  The core wavefunction was approximated with the projector augmented wave (PAW) method for Yb and Pd atoms  \cite{blochl1994,DalCorso2014}, and by the ultrasoft pseudopotential formalism \cite{Vanderbilt1990,Garrity2014} for In, Sn and H. 
The kinetic and charge density cutoffs were set to 75 and 750 Ry,
respectively \footnote{These values compare well with the convergence tests available at \href{https://www.materialscloud.org}{https://\-www.\-materials\-cloud.\-org/\-discover/\-sssp/\-table/\-efficiency} and described in Ref.~\cite{Kucukbenli2014,Lejaeghere2016}}.
The Brillouin zone integration was performed at the gamma point. The force and the total energy were minimized by using $10^{-3}$\,Ry/a.u.\ and $10^{-4}$\,Ry as thresholds, respectively. 
The candidate muon site, as resulting from DFT calculations, is shown in Fig.~\ref{fig:musite1}(b). This corresponds to four symmetry-equivalent muon-stopping 
positions, lying only 0.6\,\AA{} apart.

We also verified that the transition to a monoclinic structure (occurring at high pressure) does not modify the local magnetic field at the muon site. For this, we selected the four irreducible representations allowed by the monoclinic HP structure (space group $P2\overline{1}/c$) with nonzero magnetic moment on the Yb ion, as detailed in Table~\ref{tab:SM2}. 
Then, we calculated the dipolar field at the muon implantation sites in all these configurations, as shown in Table~\ref{tab:SM3}. It is worth noting that $B_{\mu}$ is almost unchanged for all the structures except III, where $B_\mathrm{dip}= 25.92$\,mT. 

\begin{table*} [!tbp]
	\begin{threeparttable}
		\caption{\label{tab:SM2} The four magnetic structures calculated by using the MAXMAGN code for the parent space group $P2_1/c$, labeled I to IV, that allow nonzero magnetic moments on Yb with a propagation vector (0.5,0, 0.0). Magnetic structure on the four symmetry-equivalent magnetic Yb ions in half of the unit cell, doubled along the $x$ axis.}
		\begin{ruledtabular}
			\begin{tabular}{ c @{\hspace{1em}} l @{\hspace{1.2em}} l @{\hspace{1.2em}} l @{\hspace{1.2em}} l @{\hspace{1.2em}}l} 
				& \multicolumn{5}{c}{Magnetic structure~\tnotex{tnxx59}} \\
				\cline{3-6} \\
				Label & Mag. group & Yb1 & Yb2& Yb3 &  Yb4 \\ 
				\colrule
				I   &P$_{a}$2$_{1}$/c    & ($m_x$, $m_y$, $m_z$)        & ($m_x$, $m_y$, $m_z$)          &  ($-m_x$, $m_y$, $-m_z$)        & ($-m_x$, $m_y$, $-m_z$)  \\
				II   &P$_{a}$2$_{1}$/c    & ($m_x$, $m_y$, $m_z$)    & ($-m_x$, $-m_y$, $-m_z$)     &  ($-m_x$, $m_y$, $-m_z$)  & ($m_x$, $-m_y$, $m_z$) \\
				III    &P$_{a}$2$_{1}$/c    & ($m_x$, $m_y$, $m_z$)   & ($m_x$, $m_y$, $m_z$)   & ($m_x$, $-m_y$, $m_z$) &($m_x$, $-m_y$, $m_z$)  \\
				IV  &P$_{a}$2$_{1}$/c    & ($m_x$, $m_y$, $m_z$)        & ($-m_x$, $-m_y$, $-m_z$)          &  ($m_x$, $-m_y$, $m_z$)      & ($-m_x$, $m_y$, $-m_z$)  \\
				
			\end{tabular} 
		\end{ruledtabular}
		\begin{tablenotes}
			\item[a]  \label{tnxx59}Yb1, Yb2, Yb3, Yb4 at: (0.2667, 0.66790, 0.82100), (0.2333, 0.33210,  0.17900), (0.2333, 0.1679,  0.6790), (0.2667, 0.8321, 0.32100). 
			
		\end{tablenotes} 
	\end{threeparttable}
	\
\end{table*}

\begin{table*} [!tbp]
	\begin{threeparttable}
		\caption{\label{tab:SM3}Calculated muon dipolar field $B_\mathrm{dip}$ in units of mT.~\tnotex{tnxx62}}
		\begin{ruledtabular}
			\begin{tabular}{ c @{\hspace{0.7em}} c @{\hspace{0.7em}} c @{\hspace{0.7em}} c @{\hspace{1.2em}} c @{\hspace{1.2em}}c} 
				& \multicolumn{5}{c}{$B_\mathrm{dip}$~[mT] } \\
				\cline{2-6} \\
				Label &\parbox{3cm}{$m_x=0$, $m_y=m_z$} &\parbox{3cm}{$m_x=0$, $-m_y=m_z$} &$m_x=1$, $m_y=m_z=0$ &$m_y=1$, $m_x=m_z=0$ &$m_z=1$, $m_x=m_y=0$ \\ 
				\colrule
				I  & 234.03 & 99.74 & 25.93 & 117.09 &225.85\\
				II   & 0.0 &    0.0 & 0.0  & 0.0  &0.0\\
				III    & 234.19 & 103.55 & 251.90 & 216.13 & 137.32\\
				IV  & 0.0 &    0.0 & 0.0  & 0.0 & 0.0\\
				
			\end{tabular} 
		\end{ruledtabular}
		\begin{tablenotes}
			\item[b]  \label{tnxx62} Dipolar field at the muon position (0.5, 0.5, 0.5). The substitution-induced distortions on the Yb ions were considered for the Sn/In[$x=0.5$] case. 
		\end{tablenotes} 
	\end{threeparttable}
	\
\end{table*}

\section{Averaging the dipolar fields and accounting for the effects of structural distortion}\label{app:E}

The averaging over four equivalent off-center sites leads to small 
deviations from the field value calculated at $(0.5,0.5,0.5)$. 
For completeness, in Tab.~\ref{tab:SM4}, we summarize the results 
obtained using both approaches. Only three configurations  
(here shown in bold) exhibit substantial magnetic fields at the muon 
stopping site.

\begin{table} [!tbp]
	\caption{\label{tab:SM4}Calculated dipolar field values  
		$B_\mathrm{dip}$ at the (0.5,0.5,0.5) muon-stopping site, compared 
		to those obtained after averaging over four symmetry-equivalent muon 
		positions, $\langle B_\mathrm{dip} \rangle$. The ten maximal magnetic 
		space groups of the parent $P4/mbm$ space group are here labeled I to X. 
		Dipolar fields for the Yb ions at the undistorted symmetry positions 
		(undist.) and those displaced from symmetry position induced by 
		Sn/In substitution (dist.).}
	\begin{ruledtabular}
		\begin{tabular}{ c l l l l} 
			& \multicolumn{2}{c}{$B_\mathrm{dip}$~[mT]}& &$\langle B_\mathrm{dip}\rangle$~[mT]\\
			\cline{2-3}   \cline{4-5}
			Label & undist. &  dist. & undist. & dist. \\ 
			\hline
			I   &  0.0  & 6.18 (1.91) & 0.0 & 6.37 (1.97)\\
			II   & 0.0  & 3.79 (1.66) & 0.0 & 3.86 (1.69)\\
			III  & 0.0 & 3.58 (1.76) & 0.0 & 3.71 (1.87)\\
			\textbf{IV}   & 242.16 & 245.22 (4.96) & 248.65 & 251.83 (5.20)\\
			V     & 0.0 & 4.66 (1.68) & 0.0 & 4.74 (1.71)\\
			VI    &0.0  & 6.87 (1.89) & 0.0 & 7.06 (1.95)\\
			VII   & 0.0 & 2.25 (1.21) & 0.0 & 2.36 (1.27)\\
			\textbf{VIII}  & 121.08 & 122.61 (4.53) & 124.32 & 125.91 (4.68)\\
			IX     &  0.0 & 3.03 (1.25) & 0.0 & 3.19 (1.29)\\
			\textbf{X} &  96.99 & 100.17 (4.54) & 99.26 & 102.50 (4.64)\\
		\end{tabular} 
	\end{ruledtabular}
\end{table}


\begin{thebibliography}{45}
\bibitem [{Rx()}]{Rx}%
	\BibitemOpen
	\href@noop {} {}\bibinfo {note} {In some cases we observed only tiny
		fractions of YbPd$_2$In.}\BibitemShut {Stop}%
\bibitem [{bac()}]{background}%
	\BibitemOpen
	\href@noop {} {}\bibinfo {note} {The background contribution from the Ag
		sample holder (plate no.\ 233) was determined via transverse-field
		experiments at $\mu_0H=3$\,mT and 20\,mK: $a_\mathrm{bg}\sim 0.25$ and
		$\lambda_\mathrm{bg}= 0.0158(9)$\,$\mu$s$^{-1}$.}\BibitemShut {Stop}%
\bibitem [{TF_()}]{TF_mu}%
	\BibitemOpen
	\href@noop {} {}\bibinfo {note} {A transverse-field $\mu$SR experiment at
		high temperature ($T>T_\mathrm{N}$) was used to determine $a_\mathrm{bg}$
		($\sim 65$\%). This parameter was then kept fixed during the subsequent fit
		iterations.}\BibitemShut {Stop}%
\bibitem [{ZF_()}]{ZF_mu}%
	\BibitemOpen
	\href@noop {} {}\bibinfo {note} {The K-T Gaussian depolarization rate
		$\sigma^{\mathrm{G}}_\mathrm{bg}$ (static dipolar nuclear contribution) and
		$\lambda_\mathrm{bg}$ (dynamic electronic contribution) were determined by
		fitting the ZF-asymmetry at long times, where the depolarization rate is
		dominated by the pressure cell. Both parameters were then kept fixed during
		the subsequent fit iterations except $\lambda_\mathrm{bg}$ for
		T<1K.}\BibitemShut {Stop}%
\end{thebibliography}
%
\end{document}